**Article**

# Coordinate-based neural representations for computational adaptive optics in widefield microscopy

Iksung Kang[1,†,*], Qinrong Zhang[1,6,†,*], Stella X. Yu[3], and Na Ji[1,2,4,5]

[1]*Department of Molecular and Cell Biology, University of California, Berkeley, CA 94720, USA*

[2]*Department of Physics, University of California, Berkeley, CA 94720, USA*

[3]*Department of Electrical Engineering and Computer Science, University of Michigan, Ann Arbor, MI 48109, USA*

[4]*Helen Wills Neuroscience Institute, University of California, Berkeley, CA 94720, USA*

[5]*Molecular Biophysics and Integrated Bioimaging Division, Lawrence Berkeley National Laboratory, Berkeley, CA 94720, USA*

[6]*Present address: Department of Biomedical Engineering, City University of Hong Kong, Kowloon, Hong Kong*

[†]*Equal contribution*

[*]*iksung.kang@berkeley.edu*

[*]*qzhan32@cityu.edu.hk*

# Abstract

Widefield microscopy is widely used for non-invasive imaging of biological structures at subcellular resolution. When applied to complex specimen, its image quality is degraded by sample-induced optical aberration. Adaptive optics can correct wavefront distortion and restore diffraction-limited resolution but require wavefront sensing and corrective devices, increasing system complexity and cost. Here, we describe a self-supervised machine learning algorithm, CoCoA, that performs joint wavefront estimation and three-dimensional structural information extraction from a single input 3D image stack without the need for external training dataset. We implemented CoCoA for widefield imaging of mouse brain tissues and validated its performance with direct-wavefront-sensing-based adaptive optics. Importantly, we systematically explored and quantitatively characterized the limiting factors of CoCoA's performance. Using CoCoA, we demonstrated the first *in vivo* widefield mouse brain imaging using machine-learning-based adaptive optics. Incorporating coordinate-based neural representations and a forward physics model, the self-supervised scheme of CoCoA should be applicable to microscopy modalities in general.

# Introduction

Non-invasive with subcellular resolution, optical microscopy has become an indispensable tool for biomedical research. However, sample heterogeneity and optics imperfections can introduce optical aberration and degrade image quality. Adaptive optics (AO)[1–3] techniques can be used to restore ideal imaging performance by measuring and correcting these aberrations. Conventional AO methods require specialized hardware. Direct-wavefront-sensing-based AO (DWS AO)[4–9], for example, utilizes a wavefront sensor (e.g., Shack-Hartmann sensor) for aberration measurement and a corrective device (e.g., a deformable mirror) for aberration correction, increasing the complexity and overall cost of a microscope. For indirect wavefront sensing AO methods[1–3], a corrective device is still required for wavefront correction.

Machine learning has recently emerged as a promising alternative to hardware-based AO. Supervised machine learning methods can estimate optical aberration from an experimentally measured point spread function (PSF) without the need for wavefront sensors, after a training process that involves learning a nonlinear inverse operator parametrized with neural network weights[10–18]. These approaches require an external training dataset that is either generated through simulations[10,13,14,17,18] or acquired experimentally[11,12,15]. However, to date, there is no well-established learning method for extended structures, and a corrective device is still required to correct for optical aberration for high-resolution imaging.

Here, we describe a self-supervised machine learning algorithm called CoCoA, which stands for Coordinate-based neural representations for Computational Adaptive optics, for joint estimation of wavefront aberration and three-dimensional structural recovery. Although self-supervised learning approaches have been previously used for denoising[19–23], blind deconvolution[24], 2D phase imaging[25–27], and tomography[28–31], to our knowledge, this is the first time that a self-supervised scheme is described for computational AO in fluorescence microscopy. CoCoA takes a 3D aberrated image stack as input and returns the estimated aberration and underlying structures. Representing a significant departure from the existing supervised machine learning approaches, CoCoA does not require any external supervision or external training datasets. Furthermore, CoCoA retrieves underlying features purely through computation, eliminating the need for a corrective device.

Similar to classical image deblurring problems[32–35,24], extracting wavefront and structural information from an aberrated 3D image stack is a highly ill-posed inverse problem, because there are more unknown parameters than independent measurements. To address the severe ill-posedness, CoCoA incorporated a forward model for image formation into the optimization process, obtained Zernike coefficients as a 1D vector during the optimization process, and used a multi-layer perceptron (MLP) with Fourier feature mapping (FFM) to represent complex structures. With MLPs as effective universal function approximators due to their nonlinearity[36–38] and FFM ensuring faster convergence to the optimal mapping from coordinates to structure[39,40], CoCoA carefully controlled the physical size of features reconstructed by neural networks to avoid overfitting to noise while still representing the structure accurately to an iterative non-blind baseline method[41] based on the Richardson-Lucy algorithm[42,43].

Using DWS AO to acquire the ground truth wavefront aberration, we demonstrated that CoCoA can accurately estimate aberration and retrieve 3D structural information from fixed mouse brain slices. Importantly, we characterized the performance limits of CoCoA in terms of image signal-to-noise ratio (SNR) and signal-to-background ratio (SBR). Finally, we demonstrated the first *in vivo* application of machine-learning-based AO for widefield microscopy in the mouse brain using CoCoA.

# Results

## CoCoA, self-supervised learning algorithm for computational AO

We implemented CoCoA, a self-supervised machine learning algorithm, for widefield fluorescence microscopy. CoCoA takes a single 3D image stack as input and outputs estimations of both the underlying 3D structure and the optical aberration present (**Fig. 1a**). Designed to reproduce the input image stack using a neural network model, CoCoA adjusts its parameters to identify the structure and aberration that give rise to a reproduced image stack most closely resembling the input (**Fig. 1a**). This process is referred to as *self-supervised learning*, as CoCoA learns directly from the input image stack itself without requiring labeled examples. Requiring no external supervision, CoCoA differs from existing supervised machine learning methods.

To address the challenge of representing complex structure such as neuronal processes, we employed coordinate-based neural representations[44,45,40,30] that use a MLP as a universal function approximator **(Fig. S1)**. The MLP is defined by a set of parameters denoted as $\theta$, representing the weights of the neural network. It employs FFM to achieve rapid convergence towards the optimal mapping from coordinates ($\boldsymbol{r}$) to the representation of the 3D structure ($s$). This technique allows for the incorporation of higher spatial-frequency details into the resulting representation, where the mapping can be expressed as $s = \mathcal{T}_\theta(\boldsymbol{r})$, and $\mathcal{T}_\theta$ includes a Fourier-type radial encoding scheme (**Supplementary Note**).

CoCoA also integrates a forward model for image formation into the optimization process. The model serves as a physics prior, imposing a constraint that the solution needs to satisfy the embedded mathematical model. For estimating aberration, we incorporated parameters of our microscope, including back pupil diameter, numerical aperture of the objective lens, voxel size, and emission wavelength ($\lambda$), into the forward model. We also opted to estimate the 1D-vector Zernike coefficients that represent the optical aberration at the back pupil plane.

By integrating the coordinate-based neural representations and imaging-system-informed forward model, both network-structure and physics priors were used to regularize the solution space and reduce ill-posedness. As a result, we achieved accurate estimation of both the 3D structures and optical aberration from a single image stack.

The PSF of widefield microscopy, or equivalently the image of a sub-diffraction-limit point object, is defined as

$$h = \left| \mathcal{F}\left[ G(\xi,\eta) e^{-2\pi i z \sqrt{\left(\frac{n_0}{\lambda}\right)^2 - \xi^2 - \eta^2}} \right] \right|^2 \tag{1}$$

Here, $G(\xi,\eta) = P(\xi,\eta)e^{i\varphi}$ is the complex pupil function. $P(\xi,\eta)$ describes the circular aperture of the objective lens. $\varphi$, the cumulative optical aberration at the objective lens pupil plane, equals to $\sum_{n,m} \alpha_n^m Z_n^m(\xi,\eta)$, the summation of Zernike modes $Z_n^m$ with coefficients $\alpha_n^m$ following the ANSI standard. $\mathcal{F}$ is the two-dimensional Fourier transform with respect to the pupil coordinates $\xi,\eta$. $n_0$ is the refractive index of the medium.

Given the parameterized structure $s = \mathcal{T}_\theta(\boldsymbol{r})$ and the PSF $h$, CoCoA computes the estimated 3D image stack $\hat{g}$ following the forward model for image formation as

$$\hat{g} = \mathcal{T}_\theta(\boldsymbol{r}) \ast h(\boldsymbol{r}; \alpha_n^m). \tag{2}$$

It then compares $\hat{g}$ with the experimentally acquired image stack $g$ and performs iterative updates on both the structure (via $\theta$) and the PSF (via $\alpha_n^m$) to minimize a user-defined loss function $\mathcal{L}$:

$$\mathcal{L}(\hat{g}, g; \theta, \alpha_n^m) = 1 - \mathrm{SSIM}(\hat{g}, g) + \mathcal{R}\big(\mathcal{T}_\theta(\boldsymbol{r})\big). \tag{3}$$

SSIM stands for Structural Similarity Index Metric, a widely accepted loss function[46,47,27,17], which computes the similarity between the estimated 3D image stack $\hat{g}$ and the input $g$. $\mathcal{R}\big(\mathcal{T}_\theta(\boldsymbol{r})\big)$ is a regularizer that incorporates prior information on the spatial piecewise smoothness and distribution of voxel values of the structure.

The final outputs are the estimated Zernike coefficients $\alpha_n^m$, which allows CoCoA to estimate optical aberration, as well as the neural network weights $\theta$, from which the underlying 3D structure $s$ is obtained (**Fig. 1b**). Together, the architecture of CoCoA eliminates the need for both a wavefront sensor and a corrective device. This unique joint estimation capability also sets CoCoA apart from existing supervised deep learning methods.

To characterize the performance of CoCoA, we utilized a widefield microscope equipped with an AO module composed of a wavefront sensor and a deformable mirror (DM) (**Fig. 1c**, **Fig. S2**). This system measured aberration in the emission path using DWS. With 2-photon fluorescence excitation, we generated a 3D confined 'guide star' in the sample and directed its emitted fluorescence to a Shack-Hartmann (SH) wavefront sensor after descanning[8,9]. The SH sensor used a lenslet array to segment and focus the wavefront onto a camera, creating a 2D array of foci. From local phase slopes calculated from foci displacements relative to an aberration-free condition, we were able to reconstruct the aberrated wavefront. To correct the aberration, either measured via DWS or estimated by CoCoA, we applied the opposite corrective wavefront to the DM, which modified the fluorescence wavefront before image formation on the camera. For some experiments, we also used the DM to introduce known artificial aberration to test CoCoA's performance at different imaging regimes.

One important aspect of this work was to validate the accuracy of CoCoA in aberration estimation and structural recovery, as detailed below. For aberration estimation, we used the wavefront measured by DWS as the ground truth and compared CoCoA and DWS wavefronts as well as their Zernike decompositions. For structural recovery, we compared the performance of CoCoA with the Richardson-Lucy deconvolution algorithm[42,43], a widely used computational technique, and focused on how they recovered fine neuronal features such as dendrites and dendritic spines in the brain both *in vitro* and *in vivo*.

## Implementation and two-stage learning of CoCoA

Detailed information on the neural network architecture (**Fig. S1**), hyperparameter selection (**Fig. S3**), post-processing (**Fig. S4**), and sampling (**Figs. S5,6, Table S1**) was provided in **Supplementary Note**.

From an input 3D image stack, CoCoA returns an estimated 3D structure, which, in coordinate-based neural representations, is expressed as a highly nonlinear function parameterized by MLP's weights $\theta$ with radial Fourier feature mapping[30]. In our implementation, the MLP received the radially encoded coordinates and consisted of 9 linear layers with skip connections (**Fig. S1**). In addition to the structural parameters $\theta$, for aberration estimation, CoCoA optimized the learnable coefficients $\alpha_n^m$ associated with the 17 Zernike polynomials from primary astigmatism to pentafoil, excluding defocus, following the ANSI standard.

In practice, we implemented CoCoA's self-supervised learning in two stages (**Fig. S7a**). In the first stage, we prepared a base model of the structure $\mathcal{T}_\theta(\boldsymbol{r})$ alone (i.e., without modeling the image formation process in **Eq. 2**). Starting from $\theta$ randomly chosen from a uniform distribution, we fitted the MLP network to the input image stack $g$ (normalized to have its voxel values between 0 and 1) using the loss function $\tilde{\mathcal{L}}$:

$$\tilde{\mathcal{L}}(\tilde{g}, g; \theta) = 1 - \mathrm{SSIM}(\tilde{g}, cg)\ (c > 1),\ \tilde{g} = \mathcal{T}_\theta(\boldsymbol{r}). \quad (4)$$

We utilized the Adam optimizer[48] for 400 iterations, starting with an initial rate of $10^{-2}$, and updated the learning rate using a cosine annealing learning rate schedule. At the end of Stage 1, the MLP learned network weights $\theta'$ that reproduced a scaled version of the input image stack.

In the second stage, starting with weights $\theta'$ preconditioned during the first stage, the MLP network weights were fine-tuned to generate a 3D structure and the Zernike coefficients $\alpha_n^m$ optimized, so that the 3D image stack $\hat{g}$ computed from **Eq. 2** best resembled the input image stack $g$, with $\mathcal{L}$ (**Eq. 3**) as the loss function. For the MLP network, we used an initial learning rate of $5 \times 10^{-3}$. For the Zernike coefficients, we started with $\alpha_n^m$ randomly initialized from a uniform distribution and an initial learning rate of $10^{-2}$. We employed the same learning rate schedule and optimizer as in the first stage and iteratively updated the learnable parameters by automatic differentiation on the loss function. A machine with a NVIDIA Volta 100 GPU and an Intel Xeon Gold 6248 CPU was used for computation (See **Table S2** for hyperparameter selection, experimental settings, and computation times). Code is designed and developed with PyTorch[49] and is publicly available at https://github.com/iksungk/CoCoA.

We found that CoCoA's performance benefited greatly from having the first stage of base model preparation. Starting the Stage 2 training from $\theta'$ rather than randomly initialized weights substantially reduced artifacts and improved the quality of both the structure and aberration estimation (**Fig. S7b**). Additional analysis indicated that a base model prepared from a generic fluorescent image stack can be used for Stage 2 optimization for inputs of a different sample type (**Fig. S7c-e**, **Supplementary Note**). Therefore, once a base model is available, the first stage of the two-stage learning process may be omitted for other input image stacks.

## Aberration estimation and structure recovery by CoCoA

We first tested CoCoA's performance on simulated data. CoCoA accurately extracted structures from 3D bead images of sufficient signal-to-noise ratios (**Fig. S8**). It also accurately estimated aberrations from images of single isolated beads as well as images of extended objects including 3D-distributed beads and neuronal processes (**Fig. S9**). Compared with PhaseNet, a supervised machine-learning method[14], CoCoA estimated aberration at substantially higher accuracy for all sample types but especially for complex extended objects (e.g., neuronal processes), giving confidence to its successful application to real-life images of biological samples.

We validated the efficacy of CoCoA with widefield fluorescence microscopy imaging of dendritic structures in fixed mouse brain slices (Thy1-GFP line M, **Fig. 2**). In order to introduce aberration similar to those typically induced by a glass cranial window in *in vivo* mouse brain imaging experiments[50], we placed a No. 1.5 cover glass (0.16 to 0.19 mm thickness) tilted at 3º on top of the brain slices. Before imaging, we adjusted the correction collar of the objective lens to correct for spherical aberration introduced by a 0.17-mm-thick cover glass.

We assessed the accuracy of CoCoA in estimating optical aberration through a comparative analysis of the wavefront outputs from CoCoA and from DWS. Applying the corresponding corrective wavefronts to the DM (**Fig. 2a**), we also compared their performance in improving image quality. After one round of correction, CoCoA and DWS generated similar corrective wavefronts (insets for DWS AO [1] and CoCoA [1], **Fig. 2b**) and both led to significant improvements in signal and resolution, especially for fine synaptic features (white arrowheads, insets for DWS AO [1] and CoCoA [1], **Fig. 2b**). However, CoCoA's wavefront correction resulted in slightly inferior performance compared with DWS, as indicated by the higher residual aberration (as measured by DWS after applying DWS [1] and CoCoA [1] to the DM, **Fig. 2c**) and the lower image contrast metric (bottom right, **Fig. 2c**). To further improve CoCoA's performance, we carried out iterative aberration corrections by inputting to CoCoA the 3D image stack acquired after applying the corrective wavefront from CoCoA of the previous round. Our results show that the performance of CoCoA gradually improved over three iterations, leading to comparable image quality with DWS AO (**Fig. 2b**). We also found the residual aberration after each iteration to decrease over CoCoA iterations, while DWS AO allowed diffraction-limited performance (as defined by the Rayleigh limit) after the second iteration (**Fig. 2c**).

Additionally, we evaluated the resolution improvement in the spatial frequency domain by analyzing the Fourier transform of the maximal intensity projection (MIP) image of aberration-corrected image stacks. Aberration correction, using corrective wavefront acquired through either DWS or CoCoA, led to larger magnitudes in high spatial frequency range (i.e., away from the origins in the 2D spatial frequency representations; **Fig. 2d**, left panels). The recovery of high spatial frequency information can also be easily appreciated from the radially averaged line power spectral density (PSD) profiles (**Fig. 2d**, right panels). After only one iteration, both DWS and CoCoA corrections significantly increased the power over a broad spatial frequency range when compared with 'No AO'. Compared with DWS [1], CoCoA [1] increased spectral power slightly less in the mid spatial frequency region but had similar improvement at the high spatial frequency end (inset in dashed box, **Fig. 2d**). After two iterations, CoCoA and DWS showed no perceivable difference. Quantitative Fourier Ring Correlation analyses showed similar improvements in resolution both laterally (**Fig. S10**) and axially (**Fig. S11**). These findings are consistent with residual aberration comparison and indicate that CoCoA's estimation of wavefront aberration is highly accurate.

We then investigated how the 3D structure output by CoCoA approximated the structure in real life. Because the ground-truth structural information is not available to us, we compared the structural output from CoCoA with those obtained via deconvolution, a widely applied technique that reassigns out-of-focus photons back to their sources and enhances high spatial frequency information. We applied blind and non-blind deconvolutions based on Richardson-Lucy deconvolution (RLD) algorithm[42,43] on the 'No AO' image stack used as input to CoCoA. In blind

RLD, an estimated PSF obtained from a maximum likelihood algorithm (**Methods**) was used. In non-blind RLD, the aberrated PSF from the measured aberration by DWS was directly utilized, which should lead to the most accurate deconvolution. Therefore, we used the non-blind RLD output as the standard to compare with.

Occasionally, in locations with low brightness, CoCoA encountered difficulties in accurately depicting the dim and fine features that are visible in both the DWS AO image stack and non-blind RLD structure (e.g., white arrowheads in the second row, **Fig. 2e**) or hallucinated structures that were absent from the non-blind RLD output (e.g., white arrowheads in the third row, **Fig. 2e**, also see **Supplementary Note** for relevant discussion on post-processing). Overall, however, the morphology of dendrites and dendritic spines from the CoCoA output was highly consistent with the non-blind RLD output, and the axial locations of both CoCoA and non-blind RLD outputs agree well with the two-photon fluorescence image stack (second column, **Fig. 2e**). In contrast, blind RLD reconstruction led to much noisier features, from which the sample structure cannot be ascertained with high confidence. Therefore, both being software-only algorithms, CoCoA outperformed blind RLD. Furthermore, CoCoA achieved similar performance in structural recovery to that of non-blind RLD.

## Characterizing performance limits by SNR and SBR

Although CoCoA succeeded in aberration estimation and structural recovery from the example images acquired from fixed brain slices, biological imaging often suffers from low signal-to-noise ratio (SNR, **Methods**) and signal-to-background ratio (SBR, **Fig. S12**). This is particularly true when imaging living organisms, where dim fluorophores and factors such as photodamage, photobleaching, and short exposure time (e.g., during time-lapse imaging) reduce the number of photons collected per pixel. For widefield fluorescence microscopy, larger out-of-focus fluorescence of thicker samples also leads to higher background. For all computational imaging approaches including CoCoA, images of low SNR and SBR pose challenges for their performance. Therefore, we investigated the minimum SNR and SBR thresholds required for CoCoA to be effective, before applying it for *in vivo* imaging experiments.

To control SNR, we introduced fixed amount of aberration using the DM but adjusted the post-objective power, acquiring images of increasing SNRs at higher power (**Fig. 3a**). For primary vertical coma with a 0.15λ RMS, at very low SNR values (e.g., 2.13; first column, **Fig. 3a**), there were not enough fluorescence photons to visualize features in our widefield images. Unsurprisingly, CoCoA also failed in structural recovering (first column, **Fig. 3b**). When SNR of the neuronal structures increased to 3.39, dendrites and dendritic spines could be visualized in the widefield MIP images (second column, **Fig. 3a**). However, CoCoA still failed to estimate the aberration or reveal the underlying structural features (second column, **Fig. 3b**). This was likely because even though signals of the in-focus features were sufficient for their visualization in MIP images, when out of the focal plane, the signals from these features were too noisy to be used by CoCoA for aberration estimation and structural retrieval. When the SNR increased to ~4, the performance of CoCoA markedly improved, with dendritic and synaptic features successfully retrieved (third column, **Fig. 3a,b**).

We quantified the performance of CoCoA using the Pearson correlation coefficient (PCC) and the Earth Mover's Distance (EMD)[51,52]. Using structures extracted by CoCoA from an aberration-free 3D image stack of high SNR as reference, PCC measures the correlation between CoCoA-reconstructed structures from aberrated image stacks and the reference, while EMD measures the distance between the two reconstructions by solving an optimal transport problem (**Methods**). PCC- and EMD-based quantifications confirmed the rapid performance improvement with the increase of SNR as observed by eye, with PCC increasing and EMD decreasing precipitously when SNR crosses a cutoff threshold value (**Figs. 3c,d**). Using two-segment piecewise linear fits on the PCC and EMD analysis, we found a cutoff SNR of 3.6 for this aberration, above which CoCoA provides robust structural recovery. The same cutoff also applied to aberration estimation (**Fig. 3e**). Below the cutoff, CoCoA erroneously returned non-zero coefficients for many non-primary-vertical-coma Zernike modes (gray symbols and lines, **Fig. 3e;** blue symbols and lines are the average of gray symbols and lines). Above the SNR cutoff, CoCoA accurately predicted the coefficient of primary vertical coma applied to the system (dashed black line at 0.15 λ, **Fig. 3e**) and the coefficients for the other modes were effectively zero.

Furthermore, we tested another aberration mode, primary vertical astigmatism, also at 0.15λ RMS. Using the same quantification process, we found a cutoff SNR value of 4.5 (**Figs. 3f-h**). Together, these results indicate that CoCoA performs with high accuracy when the in-focus fluorescence features had 10 ~ 20 photons per pixel by assuming a Poissonian distribution.

To experimentally control SBR levels, we introduced incrementally increasing aberration using the DM, from 0 to 0.31λ RMS (root-mean-square) by 0.04λ RMS steps. For each RMS value, we applied three different mixed-mode aberrations with randomly generated Zernike coefficients. As the aberration increased, we observed a degradation in image quality and reduction in SBR (**Fig. 4a**). Above 0.2λ RMS, the 3D neuronal structures extracted by CoCoA started to severely deviate from those acquired at higher SBR (**Fig. 4b**).

We plotted PCC and EMD against the given aberration in RMS and fitted the data points to two-segment piecewise linear curves. We carried out the same analyses for aberrations composed of low-order Zernike modes ($Z_n^m, 2 \leq n \leq 4$; primary vertical coma, astigmatism, and trefoil; **Fig. 4c-f**) or high-order modes ($n = 5$; secondary vertical coma, astigmatism, and trefoil;

**Fig. 4g-j**). For low-order aberrations, the cutoff aberration above which the reconstructed structure degraded severely was 0.19λ RMS (**Fig. 4c,d**), a value above which the wavefront estimation error increased more steeply and became larger than 0.075λ, the Rayleigh limit (**Fig. 4e**). Similarly, for aberrations containing only higher-order modes, we identified a cutoff aberration (0.16λ RMS; **Fig. 4g,h,i**) above which CoCoA gave rise to erroneous structures. The corresponding cutoff SBR for both low- and high-order aberrations was ~1.10 (**Fig. 4f,j**), indicating that CoCoA successfully retrieved structural information when the signal was 10% stronger than the background. The fact that both low- and high-order aberrations led to the same cutoff SBR suggested that the performance of CoCoA was insensitive towards the orders of the Zernike modes.

Biological samples contain features of different sizes and may vary in their fluorescence labeling density. To better understand how feature size and labeling density affect the performance limits of SNR and SBR, we carried out additional experiments on 3D tissue phantoms. These phantoms were prepared by mixing 1% agarose with fluorescent beads of either 500-nm or 2-$\mu$m diameter at varying densities. We tested phantoms with fractions of volume occupied by fluorescent beads ranging $2.35 \times 10^{-4} \sim 4.66 \times 10^{-3}$ (for 500-nm beads) and $5.11 \times 10^{-4} \sim 2.75 \times 10^{-3}$ (for 2-$\mu$m beads). We found SNR cutoffs ranging from 3.2 to 4.8 across the fluorescent volume fraction range and bead sizes (**Fig. S13**), which were consistent with the cutoffs determined from brain slices (**Fig. 3**).

We also investigated bead phantoms with fluorescence volume fraction ranging $4.96 \times 10^{-3} \sim 1.70 \times 10^{-2}$ (for 500-nm beads) and $4.78 \times 10^{-3} \sim 1.52 \times 10^{-2}$ (for 2-$\mu$m beads) (**Fig. S14**). Similar to the simulation result (**Fig. S8e-h**), the accuracy of structural retrieval, as quantified by PCC, was lowest for the densest sample (i.e., $1.70 \times 10^{-2}$ in **Fig. S14a**), likely because denser samples had images of lower SBR and more overlap between neighboring structures. However, aberration estimation accuracy and the SBR cutoff value were largely insensitive towards fluorescent volume fractions tested. For 500-nm beads, the SBR cutoff was as low as 1.03. This indicates that when SNRs are sufficiently high, CoCoA performs well even for samples with very low SBR. We also found that the 2-$\mu$m beads had higher SBR cutoff values than 500-nm beads at the similar fluorescent volume fractions. Because structural features can be considered as continuous distributions of point sources and their images composed of continuously overlapping 3D PSFs centered on these point sources, out-of-focus signals of larger features contains comparatively less information on aberration, thus requires higher SBR for aberration measurement and structural retrieval.

## CoCoA for *in vivo* imaging of the mouse brain

Having validated CoCoA for imaging fixed brain slices and investigated its performance limits, we then applied it to high-resolution *in vivo* widefield imaging through a cranial window over the left cortex of a Thy1-GFP line M mouse (**Methods**). We adjusted the correction collar of the objective lens to correct for spherical aberration introduced by the 0.17-mm-thick glass cranial window.

We first evaluated the accuracy of CoCoA in estimating optical aberration for *in vivo* mouse brain imaging by comparing its performance with DWS. Both CoCoA and DWS produced similar corrective wavefronts (**Fig. 5a**) with primary coma being the dominant Zernike mode (**Fig. 5b**), likely caused by a slight tilt of the cranial window away from being perpendicular to the optical axis of the objective. By applying the corrective wavefronts obtained from DWS and CoCoA onto the DM, we achieved higher resolution and contrast (Quantitative Fourier Ring Correlation analysis in **Fig. S15**), enabling better visualization of fine neuronal features such as dendritic spines (**Fig. 5a**, white arrowheads, and **Fig. 5c**, line signal profiles).

We next employed CoCoA to retrieve 3D neuronal structural information from the mouse brain *in vivo*. From the widefield images acquired without AO, CoCoA returned structural features such as dendritic spines that were consistent with the widefield images acquired with AO (**Figs. 5d,e**; white arrowheads). Using the same aberrated 'No AO' image stack as the input, we performed both blind and non-blind RLD (**Fig. 5e**, middle and right). Our results showed that CoCoA and non-blind RLD recovered similar synaptic structures (**Fig. 5e**, white arrowheads) while blind deconvolution resulted in artifactual structures (**Fig. 5e**, middle, red arrowheads).

The successful aberration estimation and structural recovery by CoCoA for *in vivo* imaging are to be expected, given that the SNR and SBR of the input image stacks (49.8 and 1.13, respectively, **Fig. 5d**) exceeded the cutoff values characterized previously. Notably, our experiments were conducted using illumination power within the typical range for *in vivo* widefield brain imaging experiments[53–55]. Therefore, our results indicate that CoCoA can be generally applied as a software-only approach to accurately estimate aberration and recover high-fidelity structures for *in vivo* brain imaging.

## Discussion

Utilizing coordinate-based neural representations and incorporating a physical forward model to iteratively extract structural information, CoCoA is a novel machine-learning framework that enables simultaneous wavefront aberration estimation and 3D structural recovery from a single input, an aberrated widefield image stack. A self-supervised machine learning approach, CoCoA stands apart from existing supervised machine learning methods in that it does not require an external training dataset. Recovering structural features from aberrated images, CoCoA also does not require AO hardware such as a wavefront corrective device. Moreover, we believe our physics-informed framework can easily be extended to other imaging modalities.

CoCoA is distinct from digital adaptive optics strategies that were recently developed for aberration correction and image enhancement[56,57] of two-photon synthetic aperture microscopy and scanning light-field microscopy. Although elegant and effective, these methods iteratively estimate aberration from multi-view measurements obtained either through ptychographic scanning or a lenslet array. In contrast, the standard widefield microscopy images that CoCoA utilizes are single-view images, which cannot be used for aberration estimation by these previously published methods.

Using DWS AO and RLD, we validated the performance of CoCoA in accurately estimating optical aberration and recovering structural features. Successfully demonstrating the capabilities of CoCoA in imaging neuronal structures in the living mouse brain, our work represents the first successful *in vivo* application of machine-learning-based AO for 3D structural recovery in widefield microscopy.

Importantly, we conducted a detailed investigation into the performance limits of CoCoA, specifically in terms of SBR and SNR, and determined their cutoff values required for successful CoCoA reconstruction. Our analyses suggest that there exists a fundamental lower limit on the amount of information contained in an image stack that is necessary for CoCoA to produce accurate wavefront estimation and structural information. These limits likely generally apply to all computational, including machine learning based, AO approaches.

# Methods

## Animal use

All animal experiments were conducted according to the National Institutes of Health guidelines for animal research. Procedures and protocols on mice were approved by the Institutional Animal Care and Use Committee at the University of California, Berkeley (AUP-2020-06-13343).

## AO widefield fluorescence microscope

The AO widefield microscope had two working modes (**Fig. S2**): widefield imaging mode and two-photon excitation (2PE) for AO mode. The switch between the two modes was achieved using a movable mirror (MM) controlled by an electric nanopositioning stage (SmarAct, modulator control system).

In the widefield imaging pathway (**Fig. S2a**, MM out), illumination was delivered to the sample and the emitted fluorescence was recorded by a sCMOS camera. The output beam from a 488-nm continuous laser (Coherent, Sapphire LPX 488, 400 mW) was expanded 18 times by three beam expanders (two 3×, Thorlabs, GBE03-A; one 2×, Thorlabs GBE02-A) after passing through an acoustic-optic tunable filter (AOTF; AA Opto-Electronic, AOTFnC-400.650-TN). The illumination was then relayed to the sample by three achromatic lenses (L1-L2-L3, FL = 150, 125, and 400 mm) and an objective lens (Nikon, CFI Apo LWD 25×, 1.1 NA and 2-mm WD). Emitted fluorescence was collected with the same objective. A dichroic mirror (D1, Semrock, Di-405/488/561/635-t3-25×36) was placed between L3 and the objective, reflecting illumination and transmitting collected fluorescence. The back focal plane of the objective was relayed to a DM (Iris AO, PTT489) by a pair of achromatic lenses (L4-L5, FL = 400 and 175 mm). Fluorescence reflected by the DM was then focused and imaged on a sCMOS camera (Hamamatsu, Orca Flash 4.0) by three lenses (L6-L7-L8, FL = 300, 85, and 75 mm).

In the AO 2PE pathway (**Fig. S2b**, MM in), the wavefront of a 2PE fluorescence guide was directly measured to determine artificial/sample-induced aberration. The output beam from a Ti:sappire laser (Coherent, Chameleon Ultra II) was expanded 2 times by a beam expander (2×, Thorlabs GBE02-B) after being modulated by a Pockels Cell (ConOptics, 302RM). The 2PE beam was then scanned with a pair of galvanometer mirrors (Cambridge, H2105) that are optically conjugated with a pair of achromatic lenses (L12-L11, FL = 85 mm). Another pair of achromatic lenses (L10-L9) further conjugated the galvos to the DM. For wavefront sensing, the emitted 2PE fluorescence first followed the same path as in the widefield imaging mode. The MM was placed in to reflect the fluorescence after the DM. Being relayed by L11-L12 and descanned by the galvanometer pair, the fluorescence was reflected by a dichroic mirror (D3, Semrock, Di02-R785-25×36) and relayed to an SH sensor by a pair of achromatic lenses (L13-L14, FL = 60 and 175 mm). The SH sensor was composed of a lenslet array (Advanced Microoptic Systems GmbH) conjugated to the objective back pupil plane and a camera (Hamamatsu, Orca Flash 4.0) at the focal plane of the lenslet array. Focal shifts in the SH pattern were used to calculate wavefront distortion. The corrective pattern could be then determined and applied to the DM to correct the measured aberration. When needed, 2P fluorescence imaging was enabled by placing a dichroic mirror (D1, Semrock, Di02-R785-25×36) into the light path, which reflected the emitted fluorescence to be focused on a photomultiplier tube (PMT, Hamamatsu, H7422-40). Imaging parameters can be found in **Table S2**.

## System correction and wavefront sensor calibration

In all experiments, before imaging biological samples, system aberration caused by optics imperfections and/or misalignment was measured in the widefield light path using the phase retrieval approach based on the Gerchberg-Saxton algorithm[58] from a 3D image stack of a 200-nm-diameter fluorescent bead, and corrected by the DM. Fluorescence from a 7.6×7.6 $\mu m^2$ field of 2-µm-diameter fluorescent beads (ThermoFisher Scientific FluoSpheres Carboxylate-Modified Microspheres, yellow-green 505/515) on a glass slide were 2P excited. After descanning and reflecting off the DM with system aberration correction, the recorded pattern on the SH sensor of the fluorescence wavefront was used as the aberration-free reference pattern for wavefront measurement.

## Beads sample on glass slide

The 2-μm-diameter fluorescent bead stock solution was diluted (1:500 in deionized water) and then pipetted onto a microscope glass slide precoated with poly-l-lysine hydrobromide (10 mg/ml; Sigma-Aldrich, P7890). The same method was followed to prepare 200-nm-diameter fluorescent beads sample for the validation of CoCoA in imaging sub-diffraction-limited fluorescent beads (1:10k dilution).

## Fixed mouse brain slices preparation

We prepared brain slices from a Thy1-GFP line M transgenic mouse (the Jackson laboratory, stock 007788). After being deeply anesthetized with isoflurane (Piramal), we performed a standard transcardial perfusion first with phosphate-buffered saline (PBS; Invitrogen) followed by 4% paraformaldehyde (PFA; Electron Microscopy Sciences). We then collected the mouse brain and immersed it in 2% PFA and 15% sucrose in PBS solution overnight at 4°C. After that, the immersion solution was replaced with 30% sucrose in PBS, and the brain was stored at 4°C. After another 24 hours, the mouse brain was cut to 100-μm-thick slices on a microtome (Thermo Scientific, Microm HM430). Brain slices were then placed on microscope glass slides and allowed to dry for 1 hour. Cover glass (Fisherbrand, No. 1.5) with mounting medium (Vectashield Hardset Antifade mounting medium, H-1400) was then placed on top of the glass slides with brain slices. Slices were ready for imaging after the mounting medium completely hardened.

## Cranial window implantation and *in vivo* mouse brain imaging

All Thy1-GFP line M mice (the Jackson laboratory, stock 007788) were around 4-month-old at the time of cranial window installation. The mice were deeply anesthetized under isoflurane (2.0% v/v in $O_2$) during the whole surgery. A craniotomy (3.5 mm in diameter) was created over the left cortex with dura intact. A cranial window was made by gluing (Norland 68 Optical Adhesive) together a glass ring (inner diameter: 3 mm; outer diameter: 4.5 mm) and a glass disk (diameter: 3.5 mm), both were laser cut from standard No. 1.5 microscope cover glass (Fisherbrand). The cranial window was embedded into the craniotomy and the glass ring was glued onto the skull by Vetbond (3M Vetbond). A titanium head-bar was then fixed on the skull with Vetbond and fast curing orthodontic acrylic resin (Lang Dental Mfg). *In vivo* mouse brain imaging was conducted under light isoflurane anesthesia (0.5 to 1.0% v/v in $O_2$) 2 hours after surgery.

## Calculating signal-to-background and signal-to-noise ratios

The calculation of SBR follows four steps (**Fig. S12**): (1) denoise the image stack using a 3D low-pass Gaussian kernel; (2) remove DC components and low-frequency background fluctuations using a 3D high-pass Gaussian kernel; (3) fit the image stack with a two-component Gaussian mixture model and classify the voxels into two groups (i.e., background and signal); and (4) compute SBR as the ratio of the mean of the signal voxels to the mean of the background voxels.

In order to compute SNR, we first assessed the gain of the CMOS camera to convert grayscale pixel values ($p$) to photon count ($c$) per pixel. Assuming a linear relationship between two quantities, the pixel value can be expressed as $p = \beta c$, where $\beta$ represents the gain in pixel value per photon count. Considering that $c$ follows a Poisson random distribution, with its variance $\mathrm{Var}[c]$ equal to its mean $\mathrm{E}[c]$, we derived $\beta$ to be the ratio of $\mathrm{Var}[p]$ to $\mathrm{E}[p]$. We conducted a characterization of the gain at different power levels and observed the constant gain of 2.19.

Using the signal voxels from the SBR analysis, we calculated the SNR as:

$$\mathrm{SNR} = \frac{\overline{y}/\beta}{\sqrt{\overline{y}/\beta + (n_r/\beta)^2}} \tag{5}$$

where $\beta$ is the gain in pixel values per photon count, $y$ denotes the set of signal voxels in the image stack, and $n_r$ represents the readout noise calculated as the standard deviation of pixel values in frames acquired without light exposure to the camera.

## PCC and EMD calculations

To quantify the similarity between two structural reconstructions, we employed two metrics: Pearson Correlation Coefficient (PCC) and Earth Mover's Distance (EMD). PCC is defined as the normalized inner product of the two reconstructions:

$$\mathrm{PCC}(s_1, s_2) = \frac{\sum_i (s_{1i} - \overline{s_1})(s_{2i} - \overline{s_2})}{\sqrt{\sum_i (s_{1i} - \overline{s_1})^2 \sum_i (s_{2i} - \overline{s_2})^2}}. \tag{6}$$

For EMD, we computed a Monte Carlo approximation of the *p*-sliced Wasserstein distance[59], with *p* = 2 and 200 projections used for the approximation.

## Blind and non-blind Richardson-Lucy deconvolution

Both blind and non-blind deconvolution processes were performed using a GPU-accelerated Python implementation of RLD[41]. Both non-blind and blind RLD were applied to image stacks acquired without AO correction. Non-blind RLD utilized the 3D PSF calculated from the wavefront aberration measured from DWS. The non-blind reconstructions were achieved with 500 iterations (87 sec of computation) for the brain slice sample **(Fig. 2)** and 2000 iterations (445 sec of computation) for *in vivo* mouse brain **(Fig. 5)** of RLD algorithm.

Blind RLD first estimated the 3D PSF using 100 iterations of the maximum likelihood algorithm utilizing MATLAB's deconvblind function[60], and then deconvolved the image stack with the estimated PSF using the same number of RLD iterations as the non-blind case. Consequently, blind RLD took substantially longer than non-blind RLD: estimating PSF took 1141 sec for the brain slice sample (Fig. 2) and approximately 1.25 hours for the in vivo mouse brain (Fig. 5). The deconvolution step took the same amount of time as non-blind RLD above. Therefore, the total computation time of blind RLD was 1228 sec and 1.37 hours for the slice and in vivo imaging, respectively.

## Data availability

The data used for the results in the manuscript, e.g., fixed mouse brain slice (Figure 2) and mouse brain *in vivo* (Figure 5), are available at https://github.com/iksungk/CoCoA[61]. Due to repository storage limitations, please email the corresponding authors (Iksung Kang and Qinrong Zhang) for access to the rest of the data for both the manuscript and supplementary material.

## Code availability

Code is publicly available at https://github.com/iksungk/CoCoA[61].

## Acknowledgements

This work was supported by the Weill Neurohub (N.J.) and National Institutes of Health (U01NS118300) (I.K., Q.Z., N.J.).

## Author contribution

I.K. and Q.Z. conceived of the project; N.J. supervised the project; I.K., Q.Z., and N.J. designed experiments; I.K. developed CoCoA method with inputs from S.X.Y.; Q.Z. prepared samples; Q.Z. and I.K. acquired data and prepared figures; I.K., Q.Z., and N.J. wrote the manuscript.

## Competing interests

The authors declare no competing interests.

## Ethics

All animal experiments were conducted according to the National Institutes of Health guidelines for animal research. Procedures and protocols on mice were approved by the Institutional Animal Care and Use Committee at the University of California, Berkeley (AUP-2020-06-13343).

# Figures

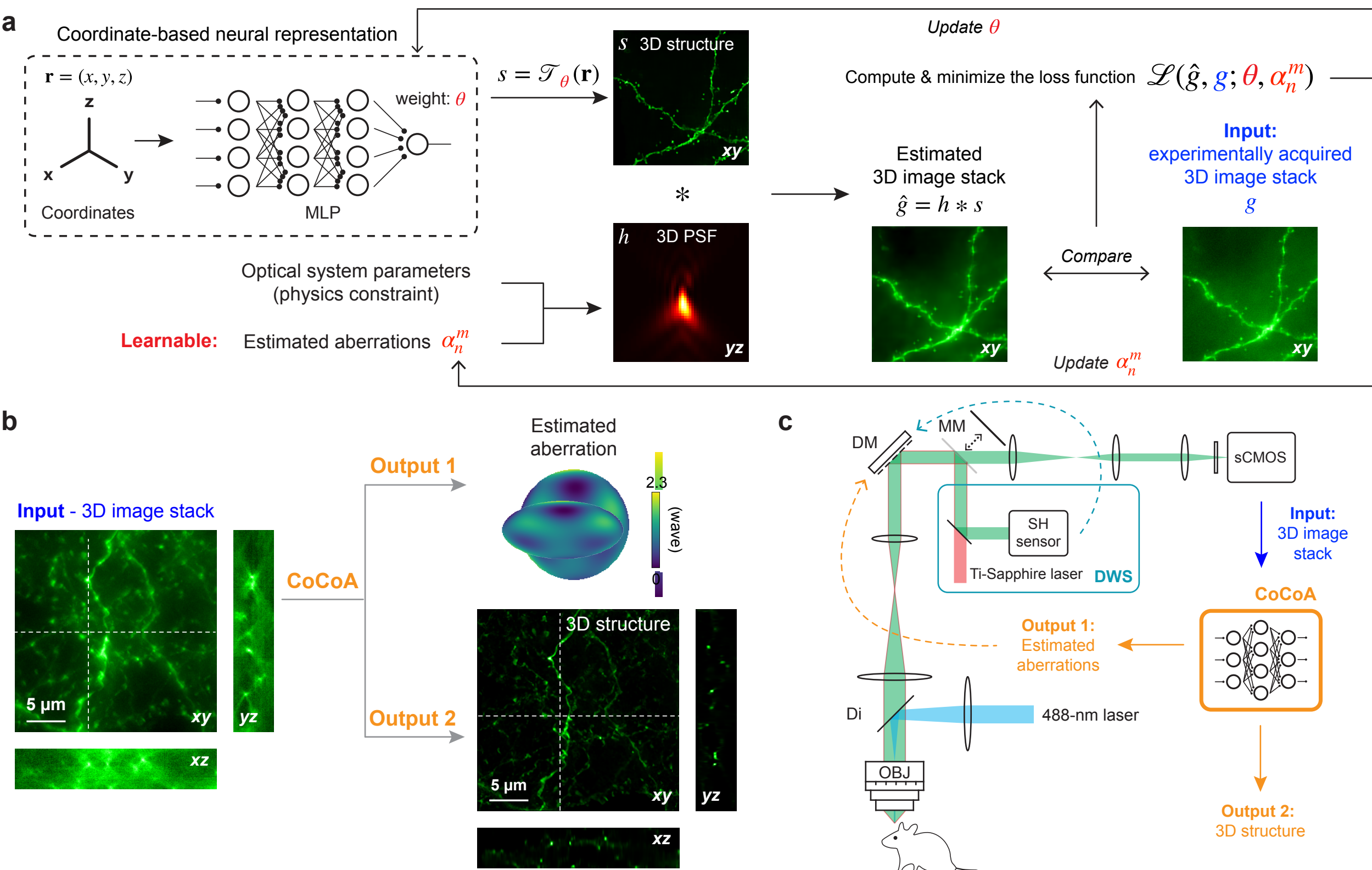

**Figure 1. Coordinate-based neural representations for computational adaptive optics (CoCoA) in widefield imaging. (a)** CoCoA's self-supervised machine learning framework iteratively updates both the 3D structure $s$, represented by a MLP with learnable weights $\theta$ through a Fourier-type radial encoding scheme $\mathcal{T}_\theta$ , and the 3D point spread function (PSF; $h$) calculated from optical system parameters and learnable Zernike coefficients $\alpha_n^m$. CoCoA minimizes a loss function $\mathcal{L}$ by comparing the image stack computed as the convolution of estimated $s$ and 3D PSF ($\hat{g}$) with the experimentally acquired 3D image stack ($g$). See **Supplementary Note** and **Fig. S1** for details. **(b)** CoCoA takes an experimentally acquired 3D image stack as input and outputs both estimated aberrations and 3D structural information. **(c)** Schematics of our widefield imaging system equipped with a Shack-Hartmann sensor (SH sensor) and a two-photon fluorescence guide star (generated by a Ti:Sapphire laser) for direct wavefront sensing (DWS) and a deformable mirror (DM) for hardware-based aberration correction. See **Fig. S2** for detailed optical path.

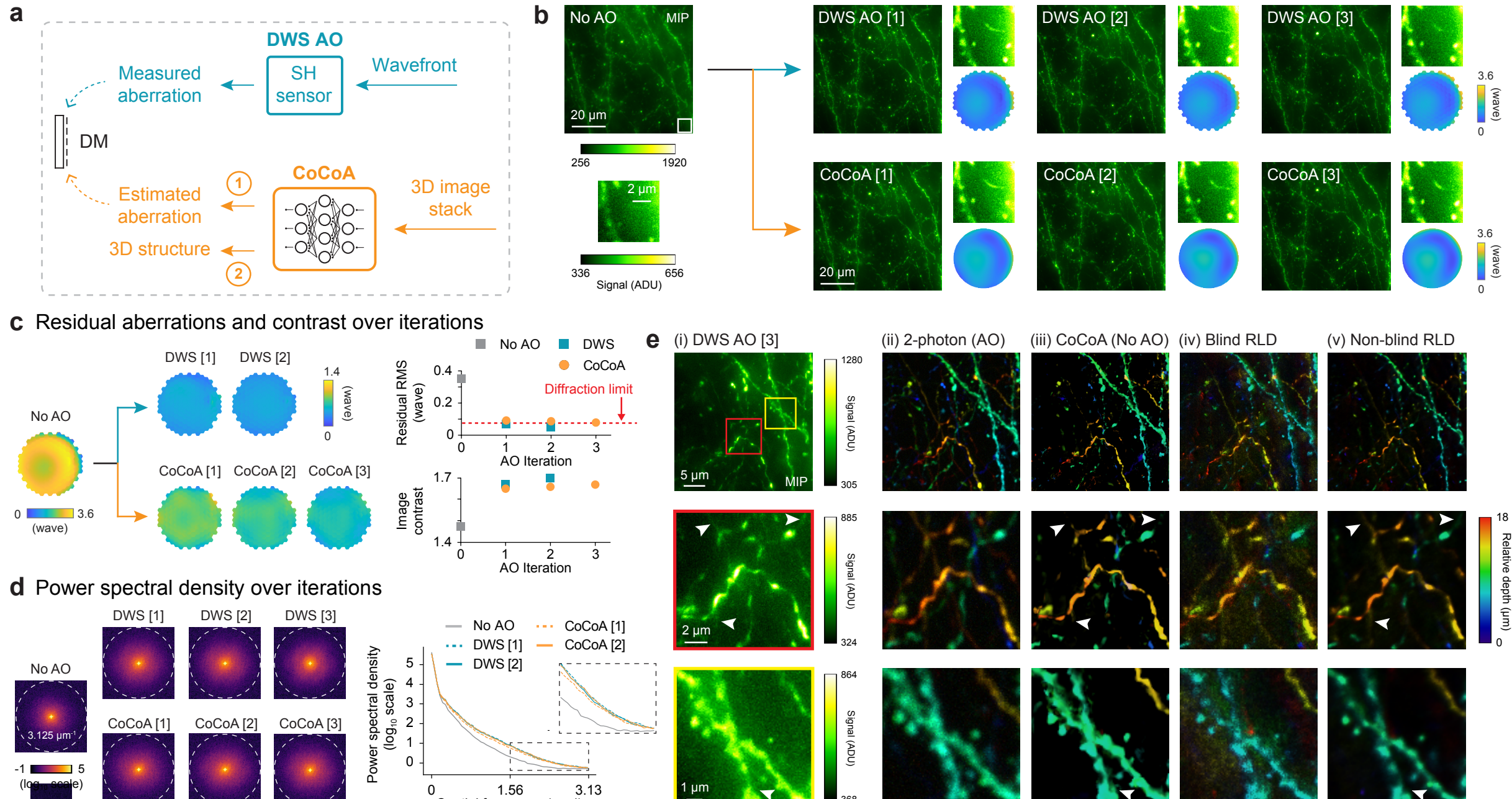

**Figure 2. CoCoA provides accurate online aberration and structure estimations as validated by DWS and non-blind Richardson-Lucy Deconvolution (RLD). (a)** Schematics of online aberration correction. Cyan: SH sensor receives a wavefront and measures wavefront aberration. Orange: CoCoA receives a 3D image stack and outputs estimated aberration and 3D structure. Corrective wavefront from either DWS or CoCoA is applied to a DM for online aberration correction. **(b)** Maximal intensity projections (MIPs) of 20-μm-thick image stacks (80 × 80 × 20 $\mu m^3$) acquired without and with aberration correction by (top) DWS and (bottom) CoCoA over iterations, respectively. Insets: zoomed-in views (white box) and corrective wavefronts. **(c)** Residual aberration and image contrast after DWS- and CoCoA-based corrections over iterations. Left: Residual aberration measured with DWS; Top right: RMS values of residual aberrations; Bottom right: image contrast computed as the ratio between the $99^{th}$ percentile and the $1^{st}$ percentile pixel values of insets in **b**. **(d)** Spatial frequency representations of images in **b** and their radially averaged profiles. Inset: zoomed-in view of a mid-to-high spatial-frequency region. Dashed circle: diffraction limit (3.125 $\mu m^{-1}$). **(e)** MIPs of image stacks (34 × 34 × 18 $\mu m^3$) acquired with (i) widefield and (ii) two-photon fluorescence microscopy after DWS AO. MIPs of reconstructed 3D structures (color-coded by depth) by (iii) CoCoA, (iv) blind RLD, and (v) non-blind RLD from No AO images.

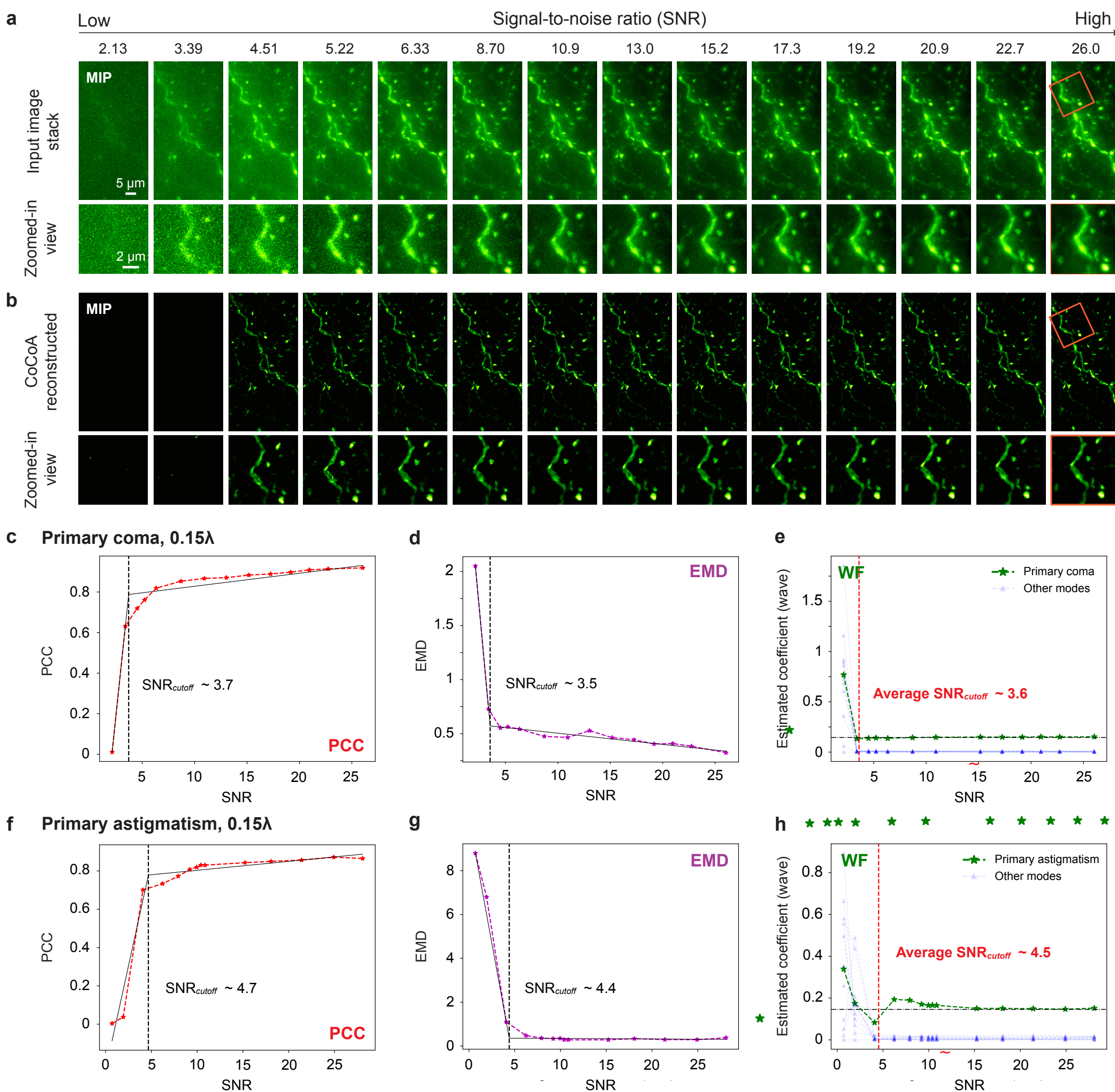

**Figure 3. CoCoA's performance depends on signal-to-noise ratio (SNR). (a)** MIPs of widefield image stacks in ascending order of SNR acquired with 0.15λ primary coma. Insets: zoomed-in views (orange box). All images individually normalized to [min, max]. **(b)** MIPs of structural stacks reconstructed by CoCoA from images in **a**, individually normalized to [min, max]. **(c)** Pearson correlation coefficient (PCC) and **(d)** Earth Mover's Distance (EMD) computed between CoCoA structure outputs from an un-aberrated image stack and aberrated input image stacks in **b**. Two-segment piecewise linear fits (solid black lines) determine SNR cutoffs (vertical dashed black lines). **(e)** CoCoA coefficients for primary coma (green symbols and lines) and other modes (gray symbols and lines) at different SNRs. Blue symbols and lines: average of other modes; Ground truth: 0.15λ for primary coma (horizontal black dash-dot line) and 0 for all other modes. Vertical red dashed line: average SNR cutoff of **c** and **d**. **(f-h)** Same as **c-e** but for 0.15λ primary astigmatism.

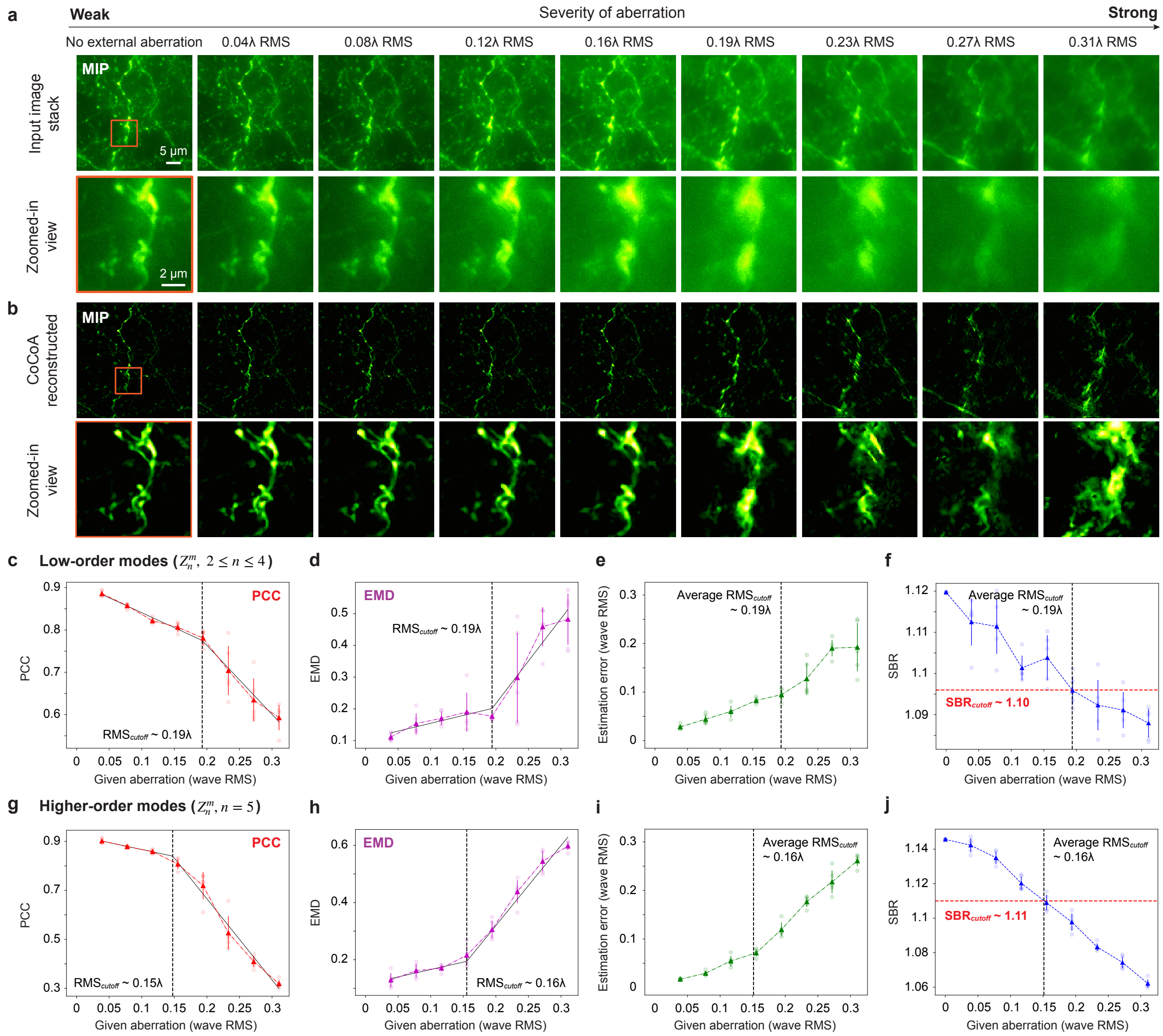

**Figure 4. CoCoA's performance depends on signal-to-background ratio (SBR).** **(a)** MIPs of widefield image stacks measured with increasingly severe aberrations (mixed low-order modes). Insets: zoomed-in views (orange box). **(b)** MIPs of structural stacks reconstructed by CoCoA from images in **a**. All MIPs in **b** individually normalized to [min, max]. **(c)** PCC and **(d)** EMD computed between CoCoA structure outputs from an un-aberrated image stack and aberrated input image stacks in **b**. Two-segment piecewise linear fits (solid black lines) determine aberration RMS cutoffs (vertical dashed black lines). **(e)** Wavefront errors in RMS between CoCoA-estimated and ground-truth wavefront aberrations. **(f)** SBR cutoff (horizontal red line) is determined from the average RMS cutoff (vertical black lines in **e** and **f**). **(g-j)** Same as in **c-f** but for mixed high-order modes. **(c-j)** Data are presented as mean values +/- SD (N= 5).

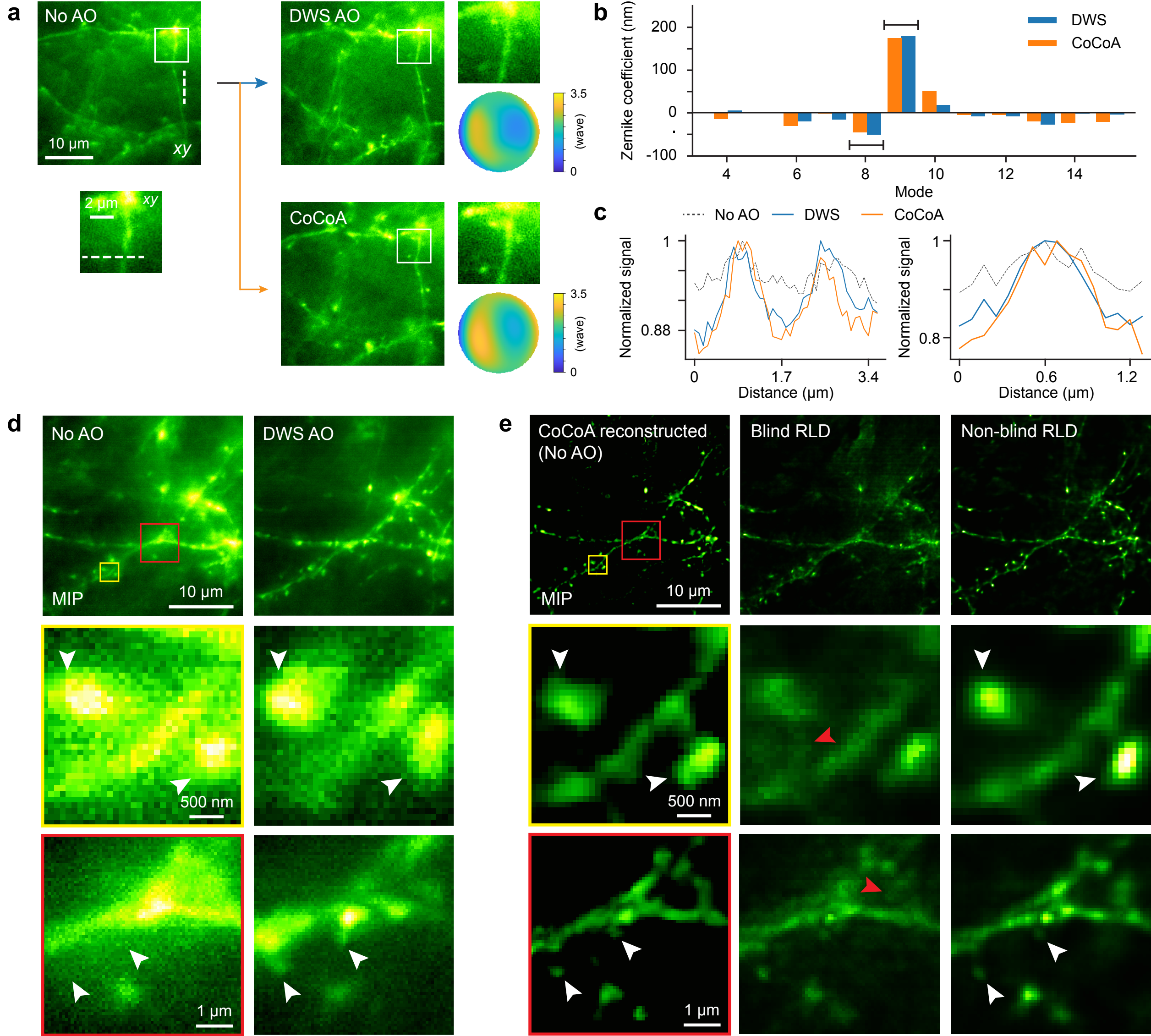

**Figure 5.** ***In vivo*** **widefield imaging of a Thy1-GFP line M mouse brain with CoCoA.** **(a)** Single widefield image planes acquired without AO, with aberration correction by DWS, and with aberration correction with CoCoA. Insets: Zoomed-in view (white box) and corrective wavefronts. All images registered using the StackReg plugin in ImageJ and then individually normalized to [min, max]. **(b)** Zernike coefficients of aberrations measured by DWS and CoCoA. Black brackets: primary coma modes. (**c**) Lateral signal profiles along dashed white lines in **a**. Line profiles are normalized by their respective maximum values. **(d)** MIPs of 4-$\mu$m-thick widefield image stacks measured without and with DWS AO and their zoomed-in views (yellow and red boxes). All images individually normalized to [min, max]. **(e)** MIPs of 3D neuronal structures reconstructed by CoCoA, blind RLD, and non-blind RLD using the 'No AO' image stack as the input. White arrowheads: synaptic and dendritic features; Red arrowheads: artifactual structures. In vivo data were acquired 10–50 $\mu$m below dura.

# Supplementary material

**Supplementary Note for Figures S1, S3 – 7**

**Figure S1.** MLP architecture in CoCoA.

**Figure S2.** Detailed optical paths of the adaptive optical widefield microscopy system.

**Figure S3.** Reconstructed structures under different hyperparameter settings in encoding, MLP, and regularization.

**Figure S4.** Post-processing suppresses spurious artifacts.

**Figure S5.** Robustness of CoCoA to downsampling image stacks along *z*-axis.

**Figure S6.** Robustness of CoCoA to downsampling image stacks along *x, y*-axes.

**Figure S7.** Two-stage training process and transfer learning performance of CoCoA.

**Figure S8.** Evaluation of structure retrieval accuracy by CoCoA from simulated 1-µm-diameter bead images.

**Figure S9.** Performance comparison of aberration estimation by CoCoA and PhaseNet from simulated image stacks of a single bead, 3D-distributed beads, and neuronal processes.

**Figure S10.** Lateral resolution estimated by Fourier Ring Correlation (FRC) for images of Thy1-GFP line M mouse brain slices.

**Figure S11.** Axial resolution improvement in images of Thy1-GFP line M mouse brain slices by DWS and CoCoA.

**Figure S12.** Overview of the procedures used to compute the signal-to-background ratio (SBR).

**Figure S13.** Determining the signal-to-noise ratio (SNR) cutoffs for images of fluorescence beads.

**Figure S14.** Determining the signal-to-background ratio (SBR) cutoffs for images of fluorescence beads.

**Figure S15.** Lateral resolution estimated by FRC for in vivo images of Thy1-GFP line M mouse brain.

**Table S1.** Effect of input image stack size on time consumption (fixed brain slice).

**Table S2.** Hyperparameter selection, experimental settings, and reconstruction / estimation times of CoCoA.

## Supplementary Note

Coordinate-based neural representation for Computational Adaptive optics (CoCoA) was designed to jointly estimate the wavefront aberration and three-dimensional structural recovery. Unlike existing supervised machine learning methods for adaptive optics (AO), CoCoA operates in a self-supervised manner, eliminating the need for external training datasets acquired either experimentally or through simulation. Instead, it only requires a single 3D image stack as input. Here and in **Figures S1, S3 – 7**, we describe our network architecture, hyperparameter selection, post-processing, generalizability, robustness to downsampling, and two-stage learning process in detail.

By employing automatic differentiation[1], CoCoA iteratively solves the inverse problem of determining both the structure $s$ and Zernike coefficients $\alpha_n^m$'s from the provided experimental 3D image stack. The neural network weights $\theta$ of CoCoA are iteratively updated to minimize the difference between the experimental image stack and the estimated stack, which is calculated as the convolution of the estimated structure and point spread function. At the end of the optimization, the estimated structure is calculated from the learned weights and the estimated wavefront aberration is calculated from the Zernike coefficients.

CoCoA's neural network architecture leverages coordinate-based learning[2–4], which entails the network learning a differentiable mapping from coordinates to structure. In CoCoA, the coordinates, initially defined in the spatial domain, are transformed to the Fourier domain using Fourier feature mapping (FFM)[2,5]. It has been demonstrated that FFM improves the performance of coordinate-based multi-layer perceptrons (MLPs) by reducing their spectral bias towards low-frequency features, which facilitates the learning of high-frequency features with faster convergence[5]. In our implementation, we chose a Fourier radial encoding scheme[3] for FFM, as written below

$$\gamma = \left(\sin\left(2^{d+1}\pi\mathcal{R}_\theta\boldsymbol{r}\right),\ \cos\left(2^{d+1}\pi\mathcal{R}_\theta\boldsymbol{r}\right)\right)_{d=0,\cdots,D,\ \ \theta=n\Delta\theta},\quad \boldsymbol{r}=(x,y,z),$$

where $\Delta\theta$ indicates the angular spacing, $d$ represents the depth of the radial encoding $\gamma$, and $\mathcal{R}_\theta$ is the rotation matrix. The angular spacing $\Delta\theta$ determines the number of spatial frequency directions, while the depth $d$ controls the maximum spatial frequency for each $k$-vector, which can be expressed as $\frac{1}{2^{d+1}\Delta x}$, with $\Delta x$ representing the $x$ and $y$ pixel size of the actual image. We observed that both $\Delta\theta$ and $d$ need to be properly chosen in order to adequately sample the Fourier space and accurately represent the structure, with smaller $\Delta\theta$ and larger $d$ required to encode more complex structures. For the structures investigated in this paper, the recommended range for the angular spacing is between 3 to 15 degrees, while the desirable range for the depth is between 6 and 7 (**Fig. S3a**).

Once the coordinates are encoded, i.e. $\gamma(\boldsymbol{r})$, they are passed through a series of linear layers (or equivalently, densely connected layers) with nonlinear activation functions (i.e., ReLU[6] in our implementation), as illustrated in **Fig. S1**. A linear layer can be understood as a collection of perceptrons, thus the multiple linear layers utilized here form an MLP. The hyperparameters of the MLP include the number of perceptrons (or features) in each linear layer and the number of linear layers. By providing a sufficient number of linear layers and enough features in each linear layer, the MLP with nonlinear activation functions gains the necessary nonlinearity and complexity to serve as a universal function approximator[7,8]. In our implementation, we found that nine linear layers, each comprising 128 features, are capable of accurately representing complex neuronal structures (**Fig. S3b**, **Table S1**).

To further regularize the space of structures represented by the Fourier-type encoding and the MLP, two regularizers are incorporated into designing our training loss function as:

$$\mathcal{R}\big(\mathcal{T}_\theta(\boldsymbol{r})\big) = \lambda\, \mathrm{TV}_z\big(\mathcal{T}_\theta(\boldsymbol{r})\big) + \kappa\, \mathrm{RSD}\big(\mathcal{T}_\theta(\boldsymbol{r})\big)^{-1},$$

which impose additional constraints on the 3D structure. $\mathrm{TV}_z(\cdot)$ is an anisotropic total-variation (TV) operator to introduce spatial piecewise smoothness in the structural features along the $z$-axis[9–11]. (TV regularization along $x$- and $y$-axes led to a large increase in computation time without significant performance improvement.) $\mathrm{RSD}(\cdot)$ computes the relative standard deviation of the structure, and its reciprocal is added to the loss function to control the distribution of voxel values of the structure[12,13]. A smaller RSD regularization parameter $\kappa$ produces a less skewed histogram profile, while a larger RSD parameter $\kappa$ results in an overly skewed histogram profile. We tested different hyperparameter settings for TV and RSD (**Fig. S3c**) and found that a suitable range for the TV regularization parameter $\lambda$ was between $10^{-9}$ and $10^{-8}$, while the RSD regularization parameter $\kappa$ should be at or below $5 \times 10^{-4}$. In our implementation, we used $10^{-9}$ for $\lambda$ and $5 \times 10^{-4}$ for $\kappa$ (**Table S1**).

Our self-supervised learning process consists of two stages (**Fig. S7a**). In the first stage, we prepared a base model and obtained network weights $\theta'$ by fitting the MLP network to an input 3D image stack using **Eq. 4** as the loss function. In the second stage, starting with the weights obtained from the first stage, CoCoA was trained to generate a 3D structure and Zernike coefficients, from which an image stack was computed to best resemble its input image stack using **Eq. 3** as the loss function. We found that starting the second stage training from $\theta'$ rather than randomly initialized weights substantially reduces artifacts and improves the quality of both the structure and aberration estimation (**Fig. S7b**).

All data in the main text utilized the same input image stack for both stages. However, we further explored the generalizability of CoCoA by performing transfer-learning tests (**Fig. S7c-e**). We compared the performance of CoCoA under both homogeneous conditions (where both stages were performed on images from the same sample type) and heterogeneous conditions (where these two stages were performed on images from different sample types). We found that the first-stage base model acquired with an input image stack of one sample type (e.g., beads) can be successfully used to train on image stacks of a different sample type (e.g., brain slices) in the second stage. The structural reconstruction and aberration estimation from both bead and neural images remained accurate, irrespective of the base model used in their second stage (**Fig. S7c-e**). Consequently, once a base model is available from bead or neural images, the first stage of the two-stage training process may be omitted for other input images.

Although CoCoA generates accurate structural estimations for the majority of neuronal structures, in local areas where signal falls below the SNR and SBR cutoffs, it can hallucinate artifactual structures. These mostly noise-related artifacts, where CoCoA misinterprets noise as structural elements, typically display hazy or speckle-like characteristics and have substantially lower voxel values than true structural components (**Figs. S4a,b**). To suppress these structural artifacts, we added a post-processing step for all images shown in the manuscript. For each image stack, we calculated an adaptive threshold of $\mu + 1.5\sigma$ of the image stack, where $\mu$ and $\sigma$ represent the mean and standard deviation of the pixel values of the stack. By setting structural voxels below the adaptive threshold to zero, this method enables us to effectively suppress hallucination and improve the quality of the reconstructed structures from experimentally acquired image stacks (**Figs. S4a,b**). For simulated images (**Fig. S8**), post-processing improved structural retrieval from image stacks of lower SNR and increased the values of the Pearson correlation coefficients (PCC) between CoCoA output and ground-truth 3D structures (**Fig. S4c,** SNR of 1.10 and 3.13), but had minimal impact on simulated bead images with high SNRs (**Fig. S4c**, SNR of 9.03 and 15.5; **Fig. S4d**) as expected. Nonetheless, there can be instances where the artifacts have intensity values comparable to real but dim structures. After post processing, residual artifacts may remain, or dim structures might be inadvertently removed (e.g., white arrowheads in **Fig. 2e**).

Our image acquisition utilized a *z* pixel size of 0.1 μm. We investigated whether data acquisition and computation time can be reduced by downsampling along the *z*-axis. We used image stacks with up to 20× downsampling (i.e., a *z*-axis pixel size of 2 μm) as inputs. We found that CoCoA maintained the quality of its structural reconstruction and aberration estimation for inputs with up to 4× downsampling, corresponding to a *z* pixel size of 0.4 μm (**Fig. S5**) and a *x, y* pixel size of 0.086 μm (**Fig. S6**). These values align with Nyquist sampling of the diffraction-limited resolution for the NA-1.1 microscope objective at 515 nm both axially (0.851 μm with Abbe's formula; Nyquist sampling: 0.426 μm pixel size) and laterally (0.234 μm with Abbe's formula; Nyquist sampling: 0.117 μm pixel size). This finding provides guidance on setting the pixel size for the input image stack. It suggests that the pixel size can be adjusted to a larger value, enabling acquisition and computation time savings, as long as it satisfies the Nyquist sampling criterion based on the ideal resolution.

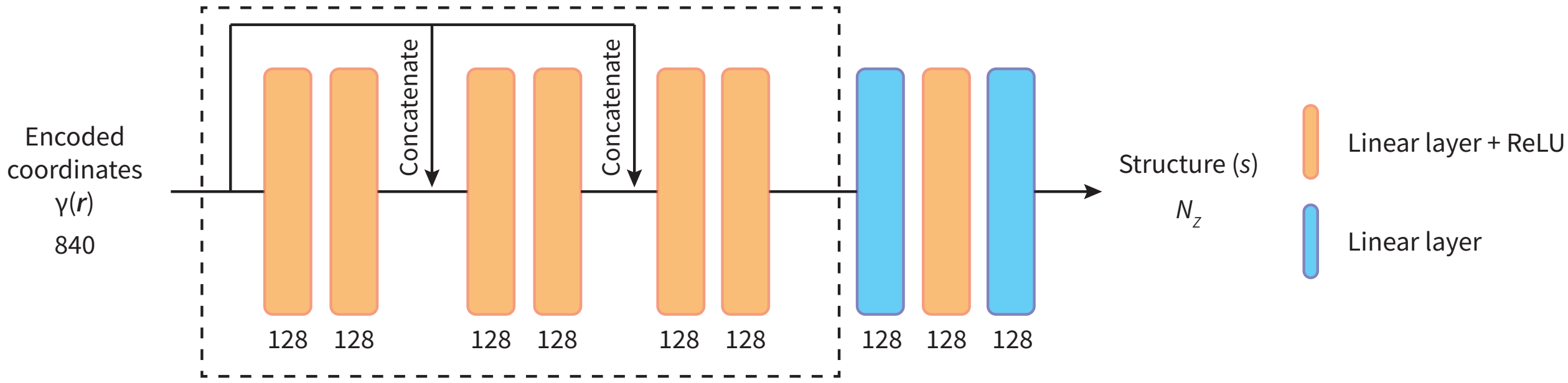

**Figure S1. MLP architecture in CoCoA.** The coordinates ($\boldsymbol{r}$) are initially defined in the spatial domain and are transformed to the Fourier domain by a radial encoding scheme. The radially encoded coordinates $\gamma(\boldsymbol{r})$ are then passed through a sequence of linear layers (or equivalently, densely connected layers) with ReLU activation functions. In our implementation, the number of features of the encoded coordinates is 840, which is directly determined by the hyperparameters of the radial encoding scheme. The numbers below the linear layers indicate their respective number of features. The neural network architecture is designed to generate the structure ($s$) as output with $N_z$ features, where each feature corresponds to one $z$-axis slice of the structure. The number of layers within the dashed box are varied in **Fig. S3b** to examine how this hyperparameter impacts CoCoA's performance. Each layer returns a tensor with the size of [$c$, $M$], where $c$ denotes the number of features in the feature axis and $M$ indicates the number of lateral pixels.

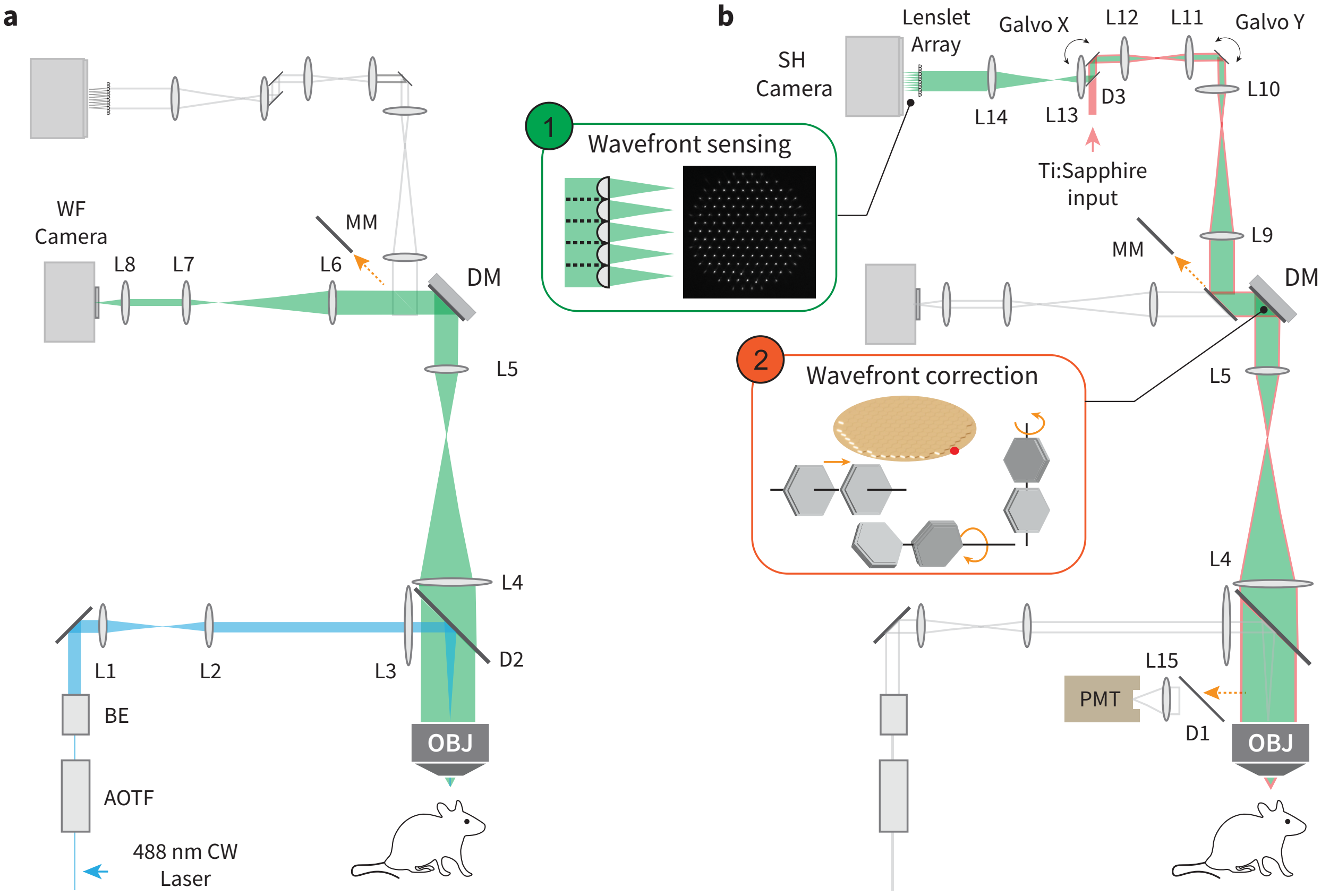

**Figure S2. Detailed optical paths of the adaptive optical widefield microscopy system. (a)** Widefield imaging light path. **(b)** Adaptive optical two-photon excitation light path. See details in **Methods**.

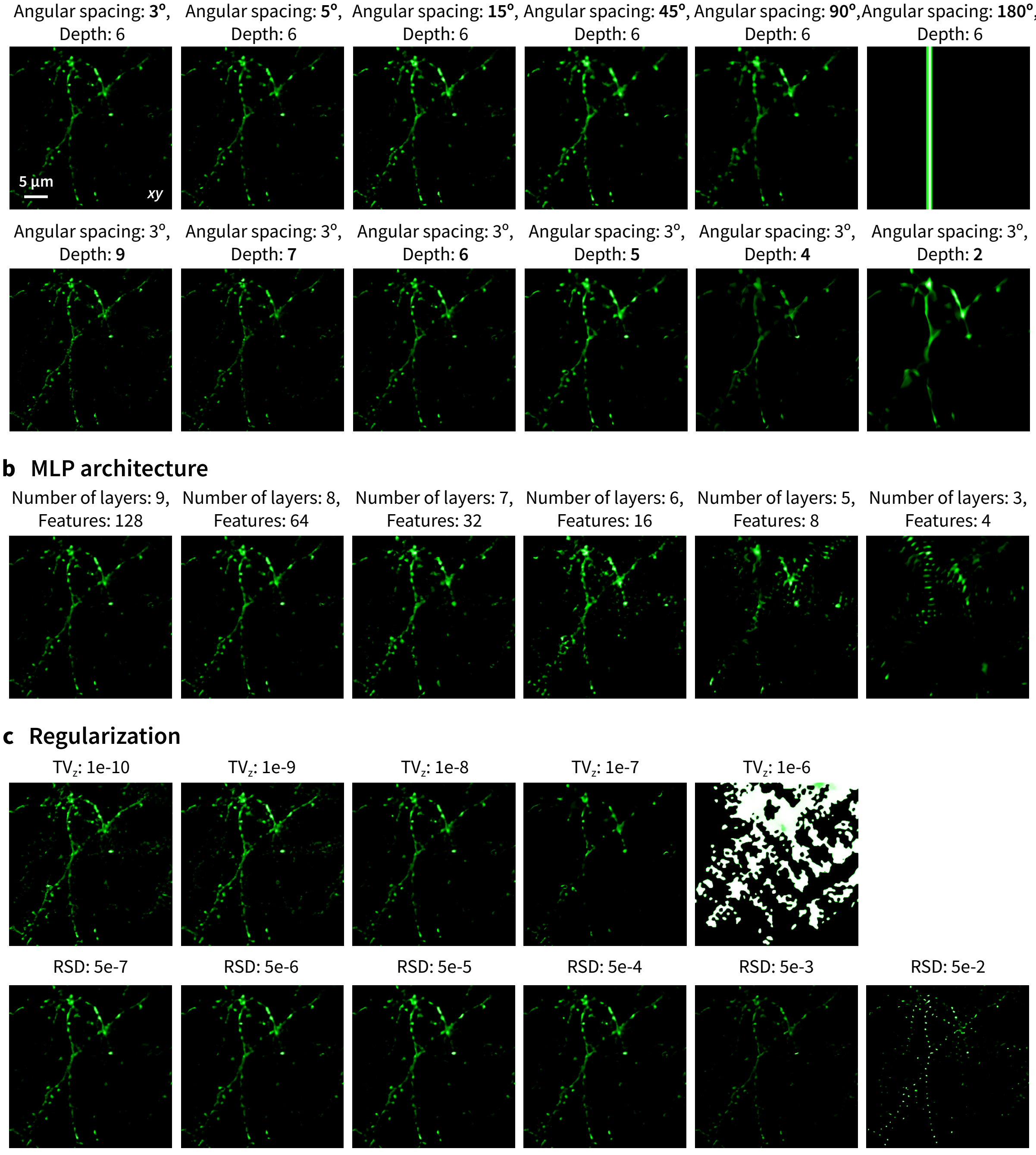

**Figure S3. Reconstructed structures under different hyperparameter settings in encoding, MLP, and regularization.** Structures recovered from a 3D image stack acquired from the mouse brain *in vivo* using various hyperparameter settings by tuning **(a)** angular spacing and depth in the radial encoding scheme, **(b)** number of layers (inside the dashed box, **Fig. S1**) and number of features of each layer, and **(c)** TV and RSD regularization parameters.

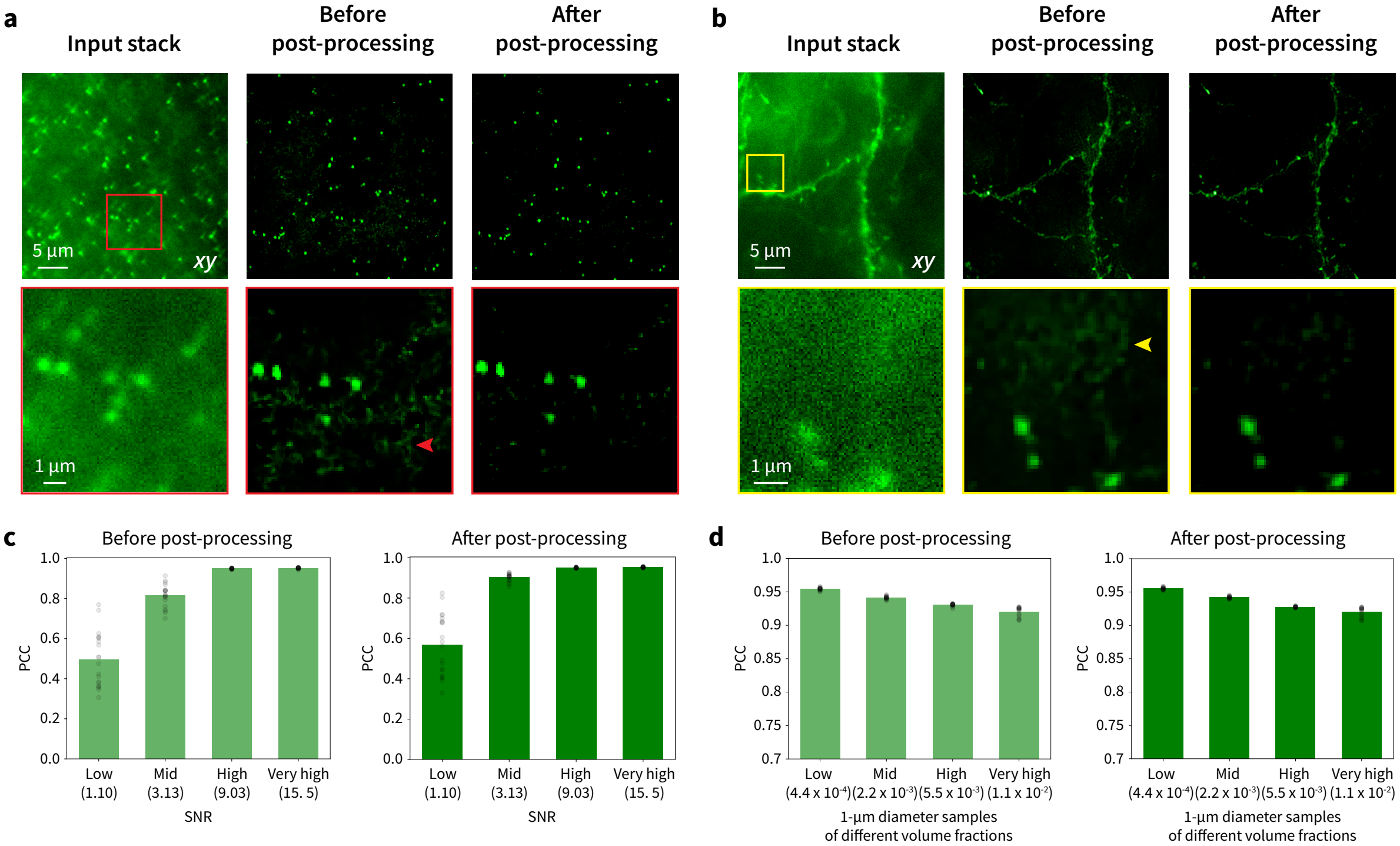

**Figure S4. Post-processing suppresses spurious artifacts.** (**a,b**) CoCoA can generate artifactual structures in dim areas of experimental images of beads and neuronal processes (zoomed-in views: red and yellow boxes, with Gamma correction $\gamma = 0.8$ to enhance the visibility of the artifactual structures indicated by arrowheads). Post-processing, by setting voxels that fall below an adaptive threshold to zero, suppresses artifacts in the reconstructed structures. (**c**) Post-processing improves structural retrieval from simulated data of lower SNR in **Fig. S8**, with Pearson correlation coefficients (PCC) between CoCoA output and ground-truth 3D structures increasing after post-processing (20 simulated datasets per SNR condition). (**d**) Post-processing has minimal impact on simulated bead images of different volume fractions in **Fig. S8** due to their high SNRs.

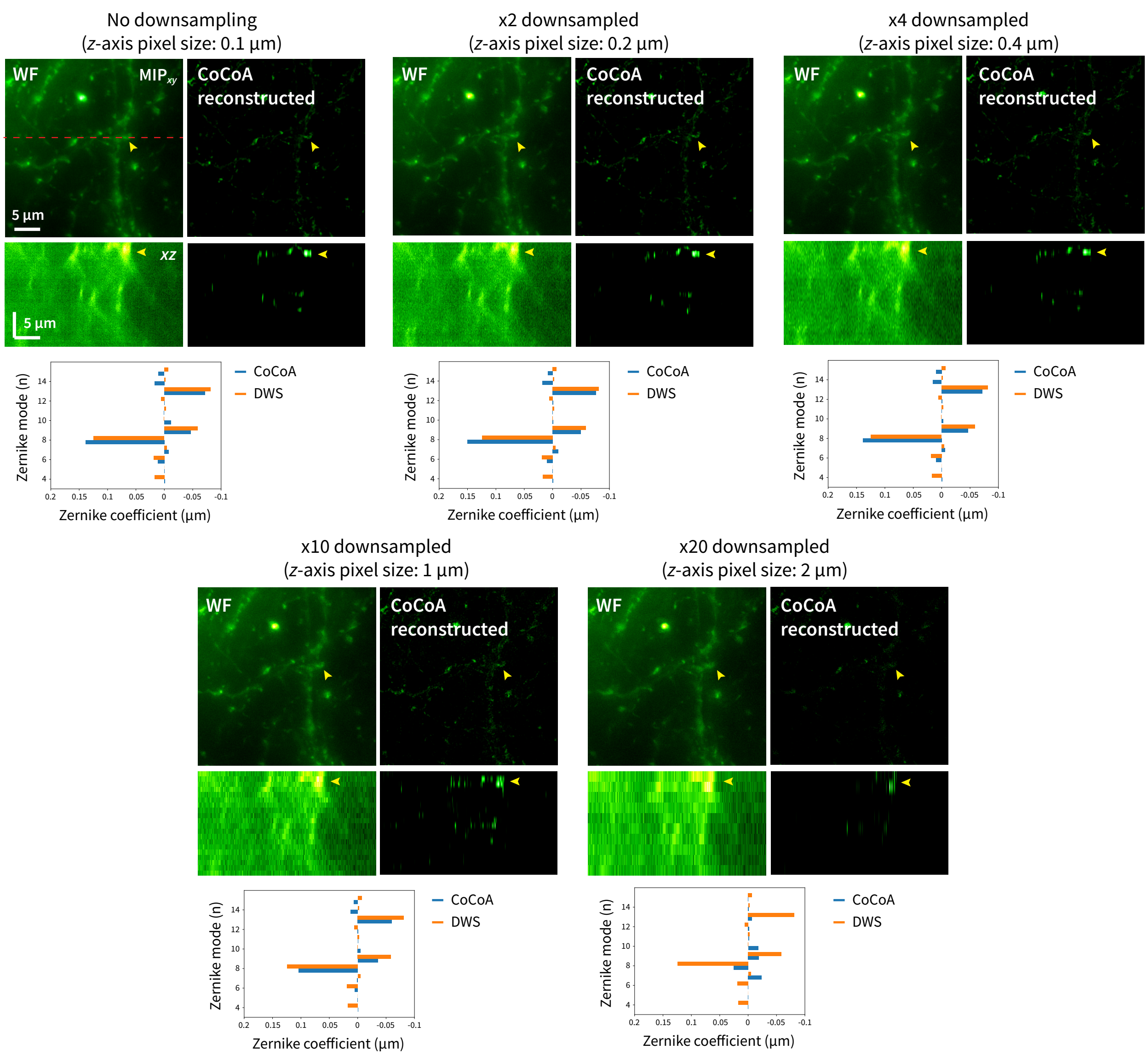

**Figure S5. Robustness of CoCoA to downsampling image stacks along *z*-axis.** Performance of CoCoA on structural and aberration estimation from input image stacks with no downsampling in *z* or 2×, 4×, 10×, 20× downsampling, respectively. Dashed red line in the MIP image indicates the line along which the *xz* image was taken. Yellow arrowheads highlight the structures of interest for comparison. The effect of downsampling along the *z*-axis on time consumption is illustrated in **Table S1**.

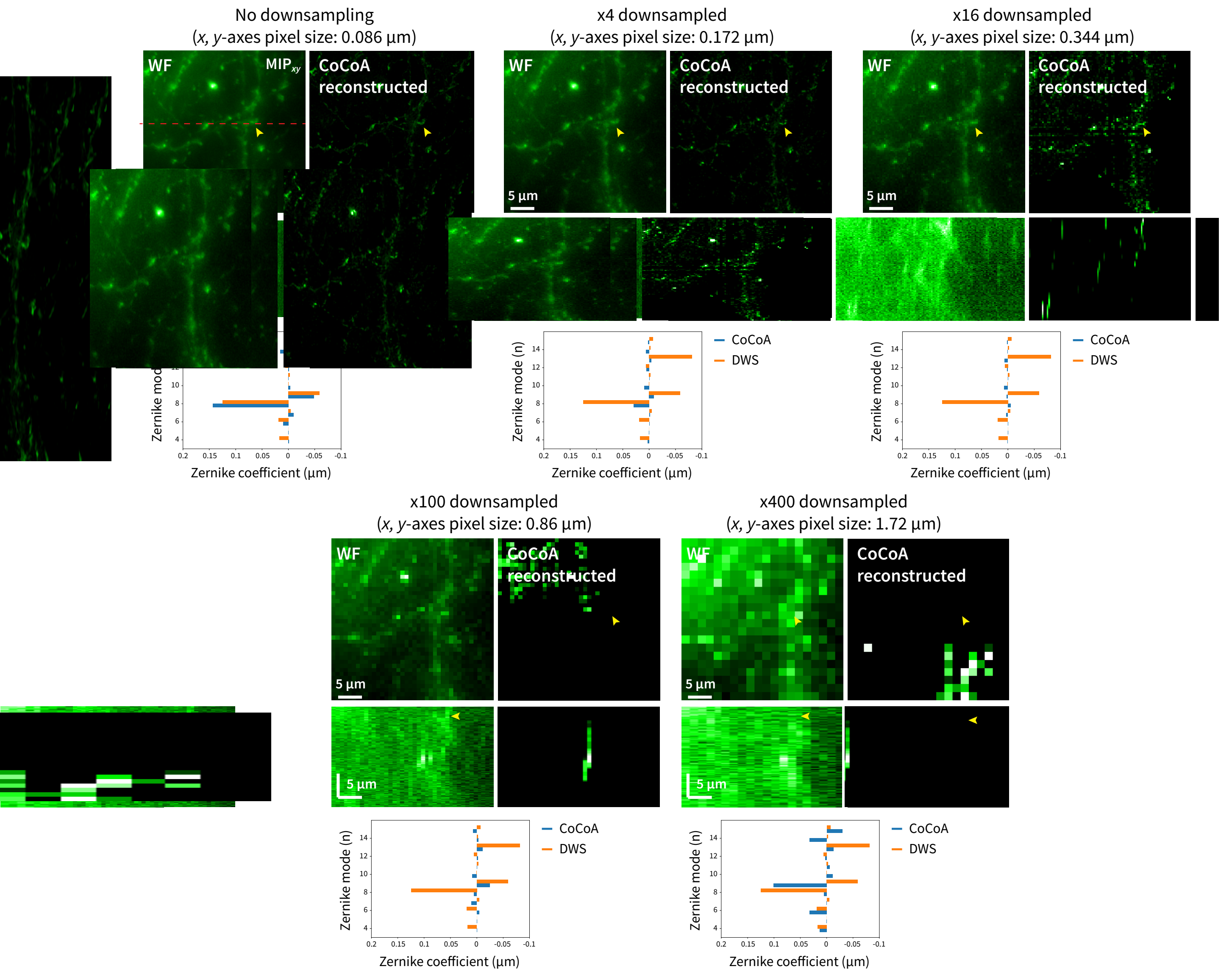

**Figure S6. Robustness of CoCoA to downsampling image stacks along *x, y*-axes.** Performance of CoCoA on structural and aberration estimation from input image stacks with no downsampling in *x, y* or 4×, 16×, 100×, 400× downsampling, respectively. Dashed red line in the MIP image indicates the line along which the *xz* image was taken. Yellow arrowheads highlight the structures of interest for comparison. The effect of downsampling along the *x, y*-axes on time consumption is illustrated in **Table S1**.

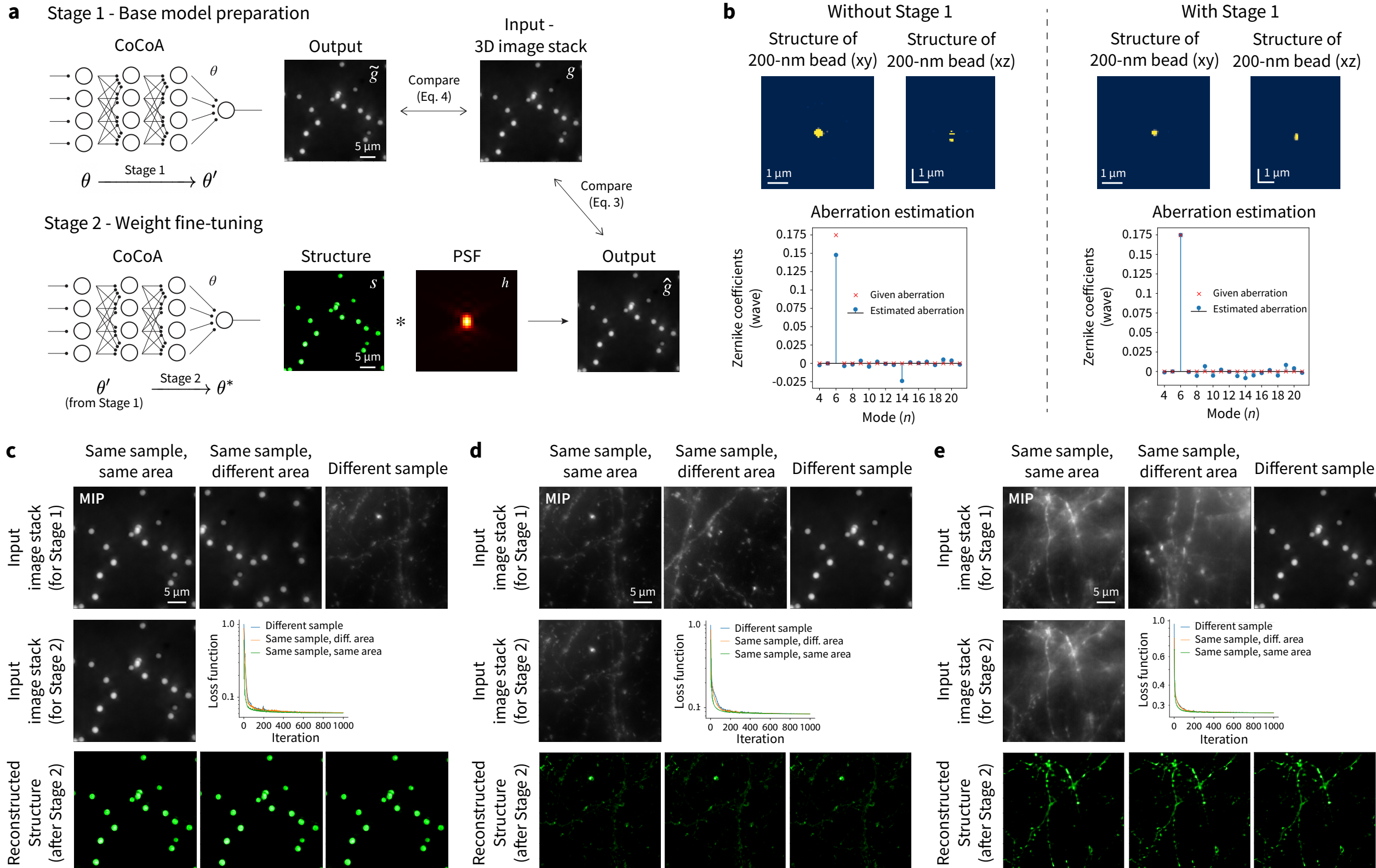

**Figure S7. Two-stage training process and transfer learning performance of CoCoA. (a)** The first stage prepares a base structure-only model with network weights $\theta'$ by fitting the network to reproduce the input image; in the second stage, the network is trained to both generate a structure by fine-tuning $\theta'$ from the first stage and estimate Zernike coefficients of aberration, so that the image stack computed from them closely resembles the input stack. **(b)** Performance comparison for structural reconstruction and aberration estimation without and with the first stage. **(c-e)** Transfer learning tests under both homogeneous conditions (where both stages were performed on images of the same sample type) and heterogeneous conditions (where Stage 1 and Stage 2 were performed on images of different sample types) reach similar performance for structural reconstruction of fluorescence beads **(c)**, brain slices **(d)**, and neurons *in vivo* **(e)**.

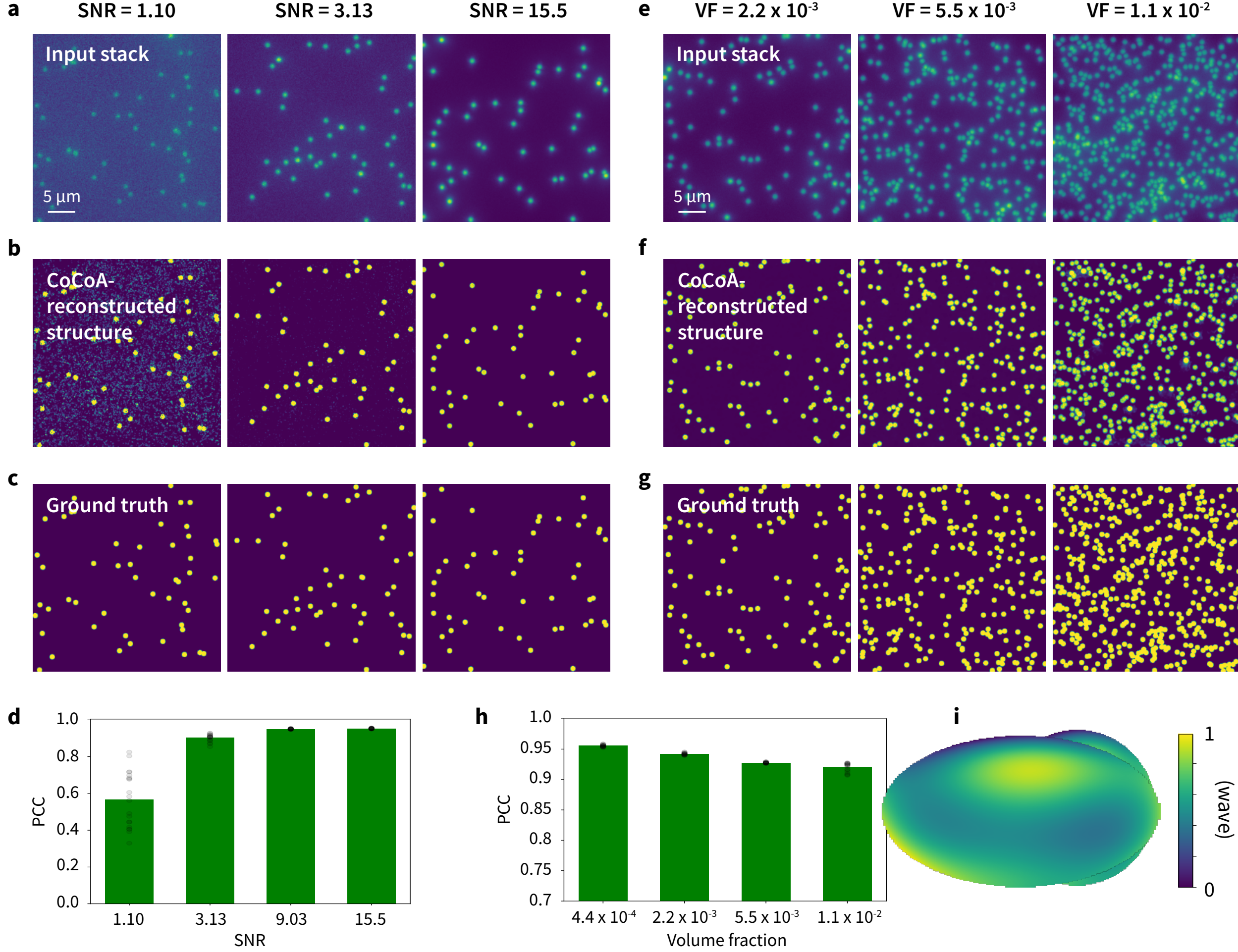

**Figure S8. Evaluation of structure retrieval accuracy by CoCoA from simulated 1-$\mu$m-diameter bead images.** **(a)** Maximal intensity projections (MIPs) of example simulated input widefield stacks (34.4 × 34.4 × 20 $\mu m^3$) of signal-to-noise ratio (SNR) of 1.10, 3,13, and 15.5, respectively. **(b)** MIPs of the 3D structural output of CoCoA. **(c)** MIP of the ground-truth 3D structures. **(d)** Pearson correlation coefficients (PCC) between CoCoA output and ground-truth 3D structures of 20 simulated datasets per SNR condition. **(e-g)** Same as **(a-c)**, except for bead samples with volume fraction (VF) of 2.2 × $10^{-3}$, 5.5 × $10^{-3}$, 1.1 × $10^{-2}$, respectively. **(h)** PCC between CoCoA output and ground-truth 3D structures of 20simulated datasets per VF value. **(i)** Ground truth aberration used in the simulation of the image stacks in **a** and **e**.

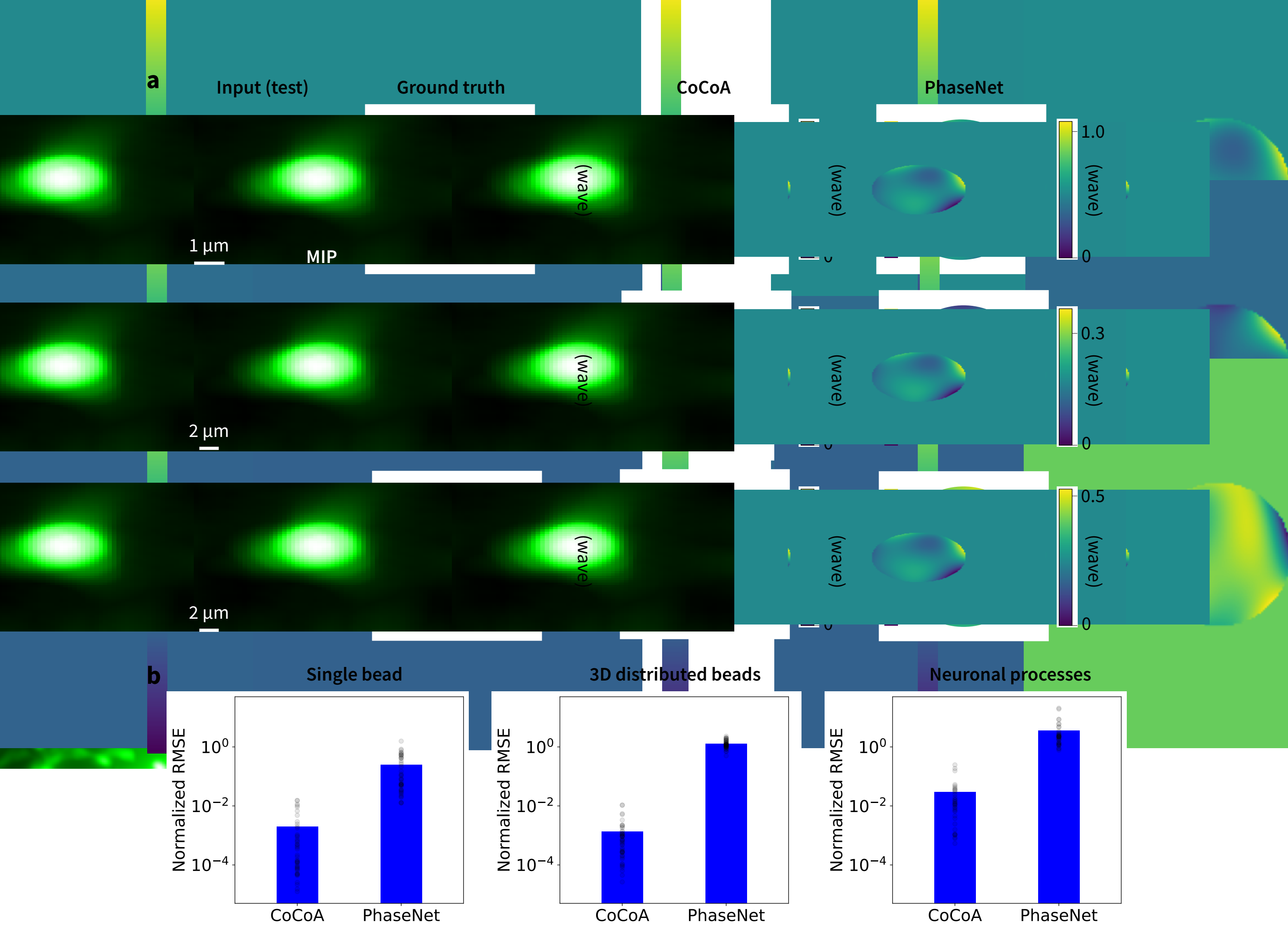

**Figure S9. Performance comparison of aberration estimation by CoCoA and PhaseNet from simulated image stacks of a single bead, 3D-distributed beads, and neuronal processes. (a)** (From left to right) MIPs of example simulated image stacks, ground-truth aberration used for simulation, aberration estimated by CoCoA, and aberration estimated by PhaseNet of (from top to bottom) an isolated single bead, 3D-distributed beads, and neuronal processes. **(b)** CoCoA and PhaseNet performance comparison for each sample type, quantified by normalized root-mean-square error (RMSE) between the estimated wavefront and ground-truth wavefront (20 simulated datasets per sample type).

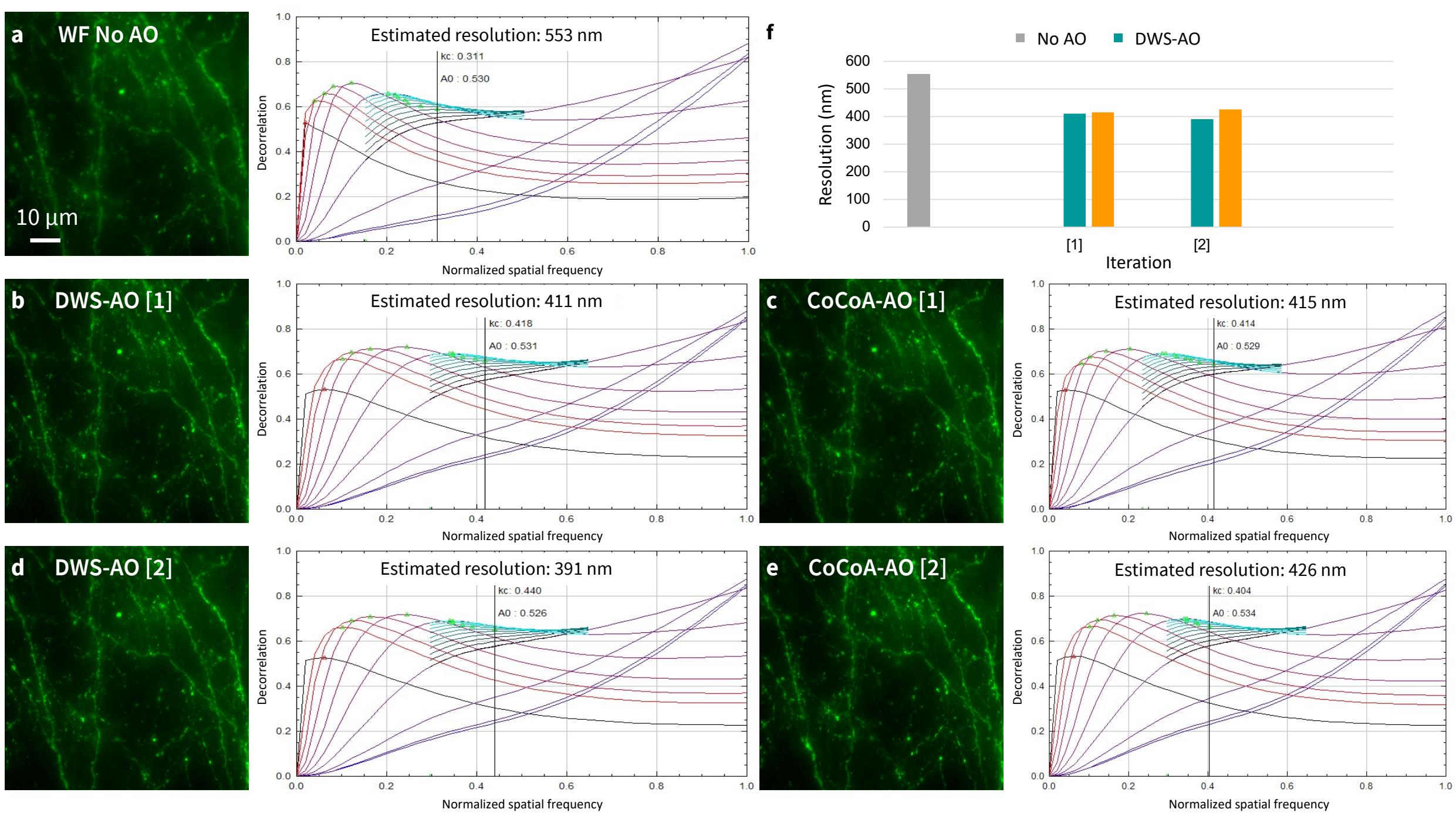

**Figure S10. Lateral resolution estimated by Fourier Ring Correlation (FRC) for images of Thy1-GFP line M mouse brain slices. (a-e)** Left: Maximum intensity projections (MIPs) of widefield image stacks ($34 \times 34 \times 18$ $\mu m^3$) acquired **(a)** without and with aberration correction by **(b,d)** DWS and **(c,e)** CoCoA over two iterations, respectively. Right: corresponding FRC output plots with estimated resolutions. (**f**) Estimated resolutions without and with DWS- and CoCoA-based corrections over iterations. Images shown are the same as in **Fig. 2b**. All images were individually normalized.

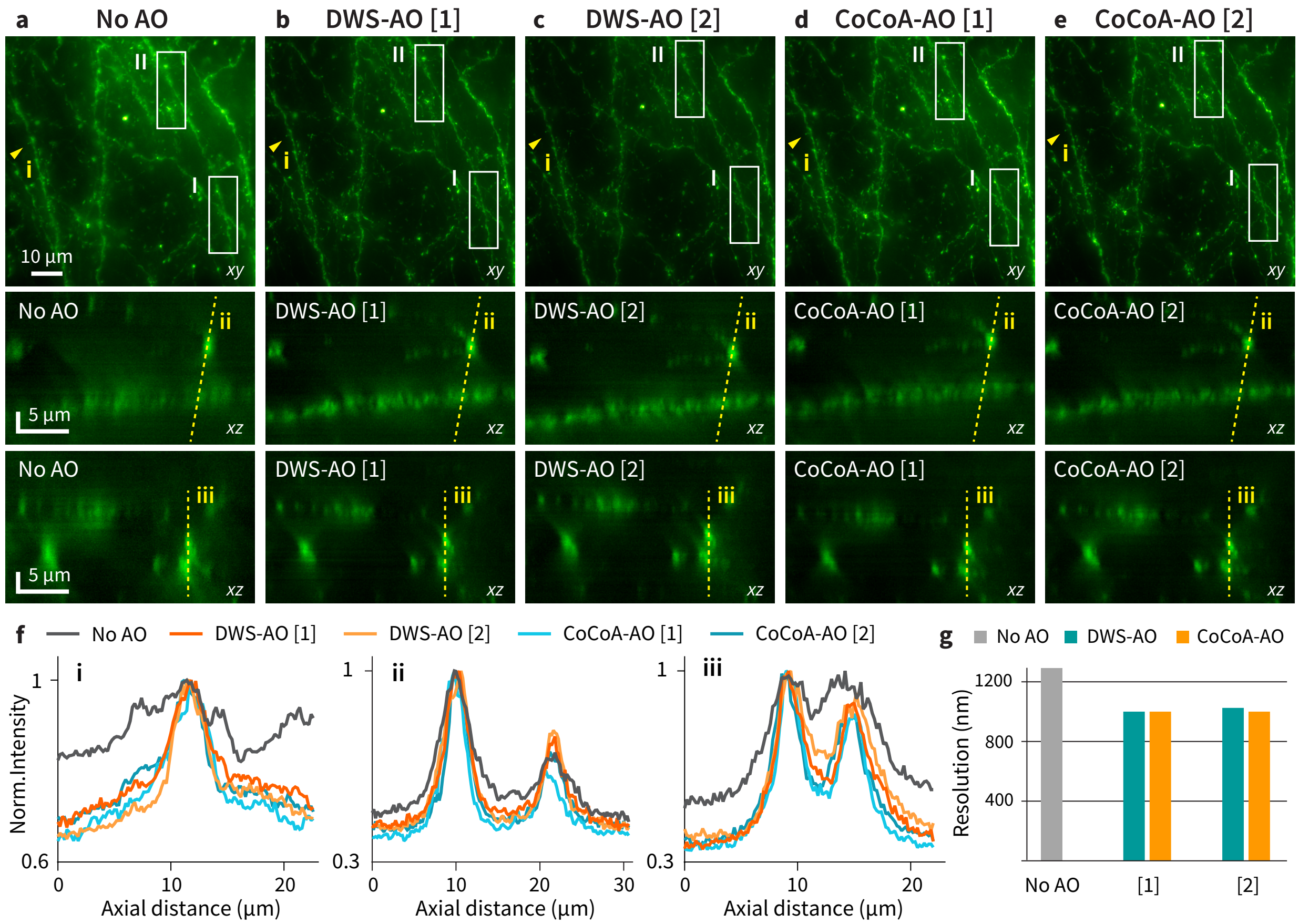

**Figure S11. Axial resolution improvement in images of Thy1-GFP line M mouse brain slices by DWS and CoCoA. (a-e)** Top: Maximum intensity projections (MIPs) of widefield image stacks (34 × 34 × 18 $\mu$m$^3$) acquired **(a)** without and with aberration correction by **(b,c)** DWS and **(d,e)** CoCoA over two iterations, respectively. Bottom: axial MIPs along the y axis from white boxes I and II. **(f)** Axial signal profiles of a dendritic spine i (yellow arrowhead), and of dendritic structures along the yellow dashed lines ii and iii. **(g)** Axial resolution estimated using FRC, where a 1/8-sectorial mask was applied. All images were individually normalized with contrast adjusted for better visualization.

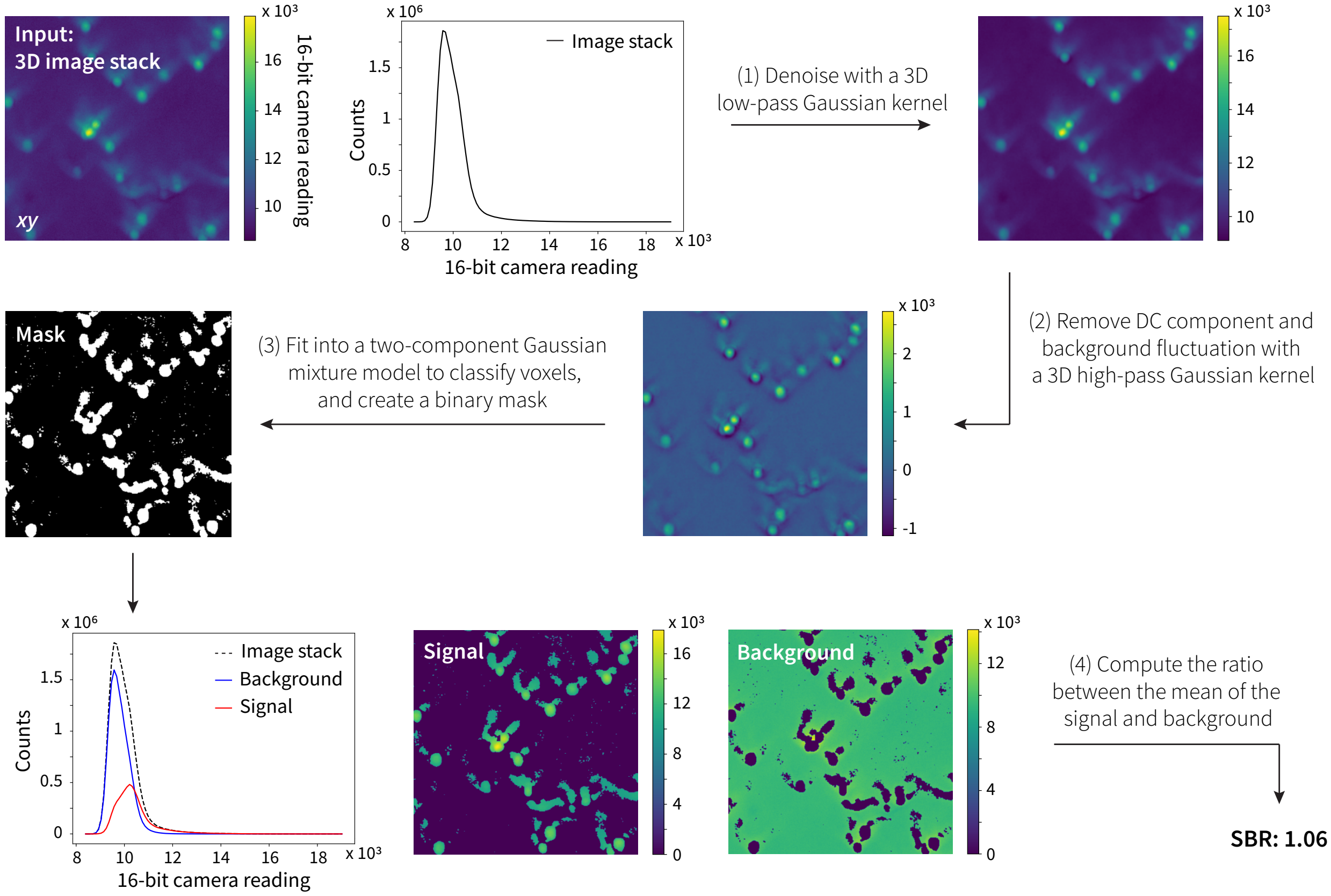

**Figure S12. Overview of the procedures used to compute the signal-to-background ratio (SBR).** Additional details and code can be found in https://github.com/iksungk/CoCoA.

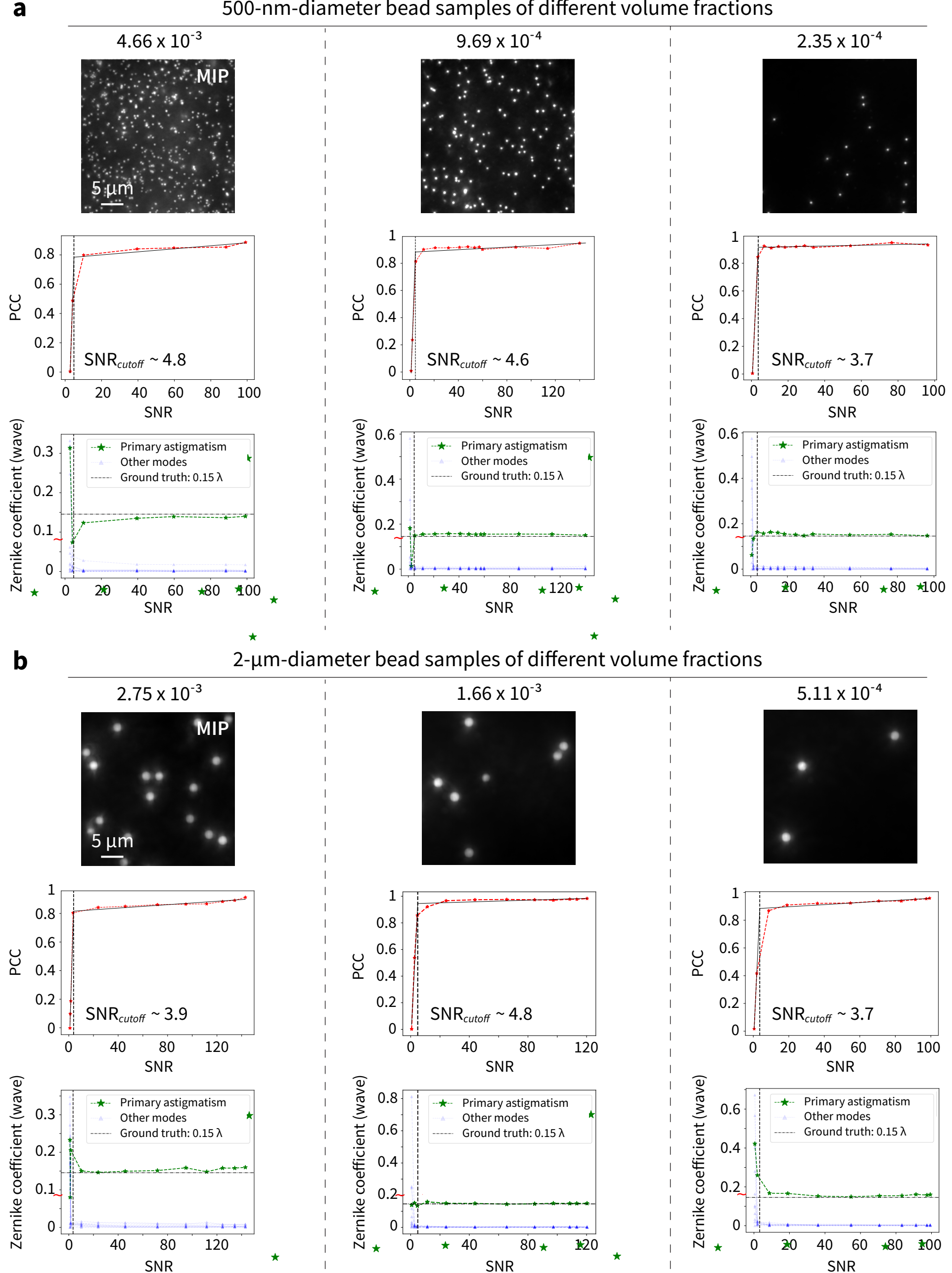

**Figure S13. Determining the signal-to-noise ratio (SNR) cutoffs for images of fluorescence beads.** Image stacks of bead samples with different volume fractions and diameters (**a**: 500 nm; **b**: 2 µm), acquired at different SNRs, were used as inputs. For all datasets, 75 nm RMS (equivalent to 0.15λ for Zernike coefficient) of primary astigmatism was introduced into the widefield microscope using a deformable mirror. For each column: (Top) Maximal intensity projections (MIP) of an example input bead image stack; (middle) Pearson correlation coefficients (PCC) between the structure outputs reconstructed from un-aberrated image stacks and aberrated image stacks of different SNRs; (bottom) Comparison of the estimated wavefront aberration for different aberration modes by CoCoA with the ground-truth aberration (0.15λ for primary astigmatism; 0 for all other modes). SNR cutoffs (vertical black dashed lines) were determined by two-segment piecewise linear fits on the PCC analysis.

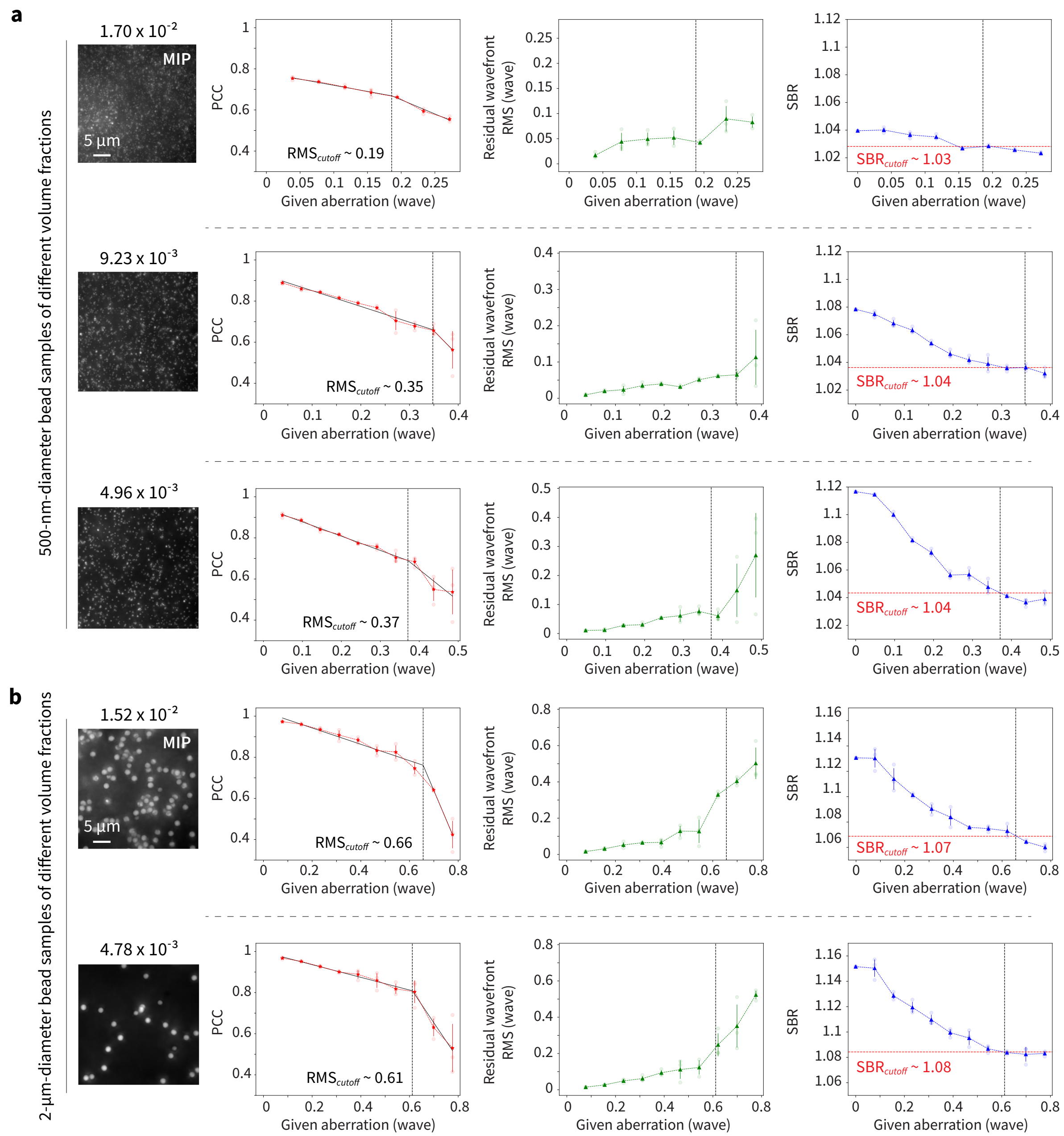

**Figure S14. Determining the signal-to-background ratio (SBR) cutoffs for images of fluorescence beads.** Image stacks of bead samples with different volume fractions and diameters (**a**: 500 nm; **b**: 2 μm), acquired under increasing aberration, were used as inputs. (Aberration with mixed low-order modes was introduced into the widefield microscope using a deformable mirror.) For each row (from left to right): MIP of an example input bead image stack; Pearson correlation coefficients (PCC) between the structure output without external aberration and the structure output with given aberrations; Residual wavefront error between CoCoA estimation and ground-truth aberration; SBR of the input image stack versus given aberration. RMS cutoffs (vertical black dashed lines) were determined by two-segment piecewise linear fits on the PCC analysis; SBR cutoffs (horizontal red dashed lines) were read from the SBR-versus-given-aberration plots with the given aberration being at $RMS_{cutoff}$. **(a-b)** Data are presented as mean values +/- SD (N = 3).

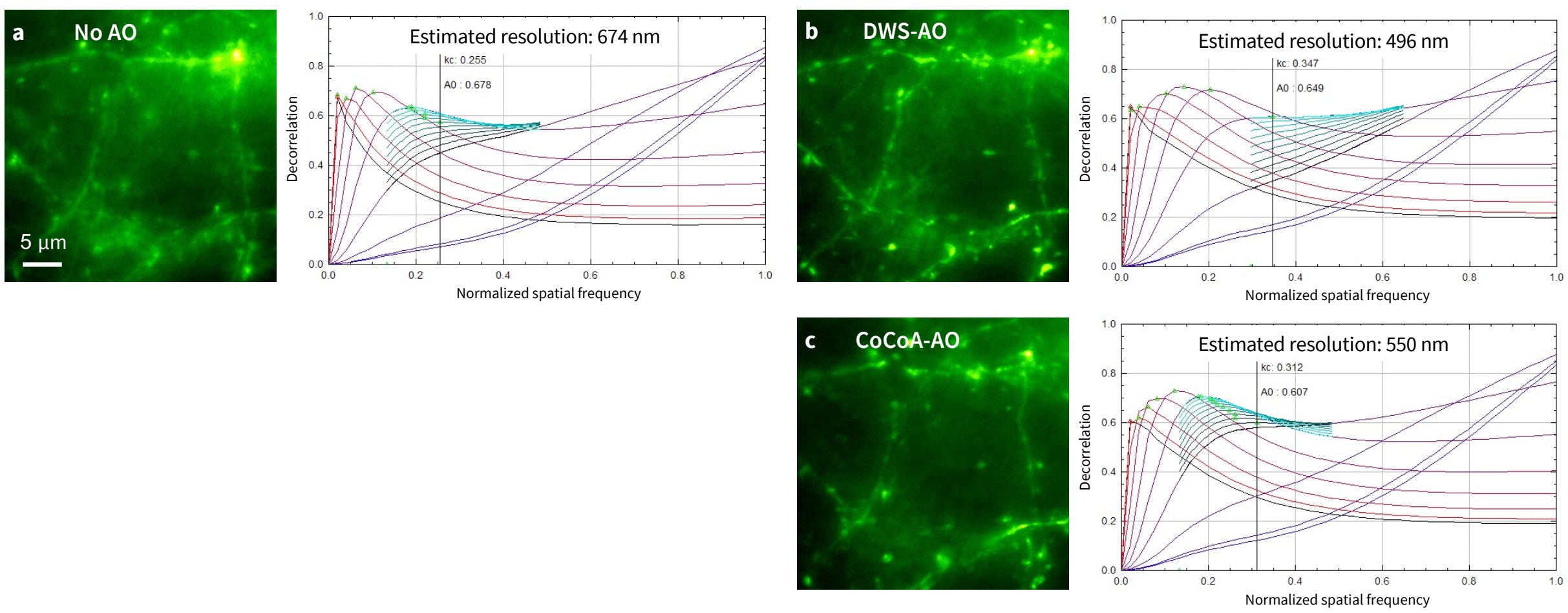

**Figure S15. Lateral resolution estimated by FRC for in vivo images of Thy1-GFP line M mouse brain.** **(a-c)** Left: MIPs of widefield image stacks ($34 \times 34 \times 12\ \mu m^3$) acquired **(a)** without and with aberration correction by **(b)** DWS and **(c)** CoCoA. Right: corresponding FRC output plots with estimated resolutions. Same dataset as in **Fig. 5a**. All images were individually normalized.

**Table S1. Effect of input image stack size on time consumption (fixed brain slice)**

| Downsampling along the $z$-axis | | | | |
|---|---|---|---|---|
| **Downsampling factor ($z$ pixel size)** | **Number of voxels $[z, y, x]$ (px)** | **Duration of Stage 1 (s)** | **Duration of Stage 2 (s)** | **Total duration (s)** |
| 1 (0.1 $\mu$m) | [200, 400, 400] | 28.84 | 304.0 | 332.8 |
| 2 (0.2 $\mu$m) | [100, 400, 400] | 24.11 | 193.2 | 217.3 |
| 4 (0.4 $\mu$m) | [50, 400, 400] | 22.30 | 147.8 | 170.1 |
| 10 (1.0 $\mu$m) | [20, 400, 400] | 21.64 | 121.0 | 142.6 |
| 20 (2.0 $\mu$m) | [10, 400, 400] | 21.25 | 116.1 | 137.4 |

| Downsampling along the $x, y$-axes | | | | |
|---|---|---|---|---|
| **Downsampling factor ($x, y$ pixel size)** | **Number of voxels $[z, y, x]$ (px)** | **Duration of Stage 1 (s)** | **Duration of Stage 2 (s)** | **Total duration (s)** |
| 1 (0.086 $\mu$m) | [100, 400, 400] | 24.11 | 193.2 | 217.3 |
| 4 (0.172 $\mu$m) | [100, 200, 200] | 5.410 | 47.07 | 52.48 |
| 16 (0.334 $\mu$m) | [100, 100, 100] | 1.641 | 15.92 | 17.56 |
| 100 (0.86 $\mu$m) | [100, 40, 40] | 1.344 | 14.64 | 15.98 |
| 400 (1.72 $\mu$m) | [100, 20, 20] | 1.300 | 14.39 | 15.69 |

## Table S2. Hyperparameter selection, experimental settings, and reconstruction / estimation times of CoCoA.

| Hyperparameters | Fig. 2E | Fig. 3 | Fig. 4 | Fig. 5D | Fig. S4A | Fig. S4B | Fig. S5A | Fig. S5B |
|---|---|---|---|---|---|---|---|---|
| Sample type | Fixed mouse brain slice (Thy1-GFP line M) | Fixed mouse brain slice (Thy1-GFP line M) | Fixed mouse brain slice (Thy1-GFP line M) | Thy1-GFP line M mouse brain *in vivo* | 500-nm diameter fluorescent beads | 2-μm diameter fluorescent beads | 500-nm diameter fluorescent beads | 2-μm diameter fluorescent beads |
| Dimensions of input image stack | $34 \times 34 \times 20$ μm$^3$ | | | $34 \times 34 \times 40$ μm$^3$ | $34 \times 34 \times 20$ μm$^3$ | | | |
| Pixel size (*dx, dy, dz*) | (86 nm, 86 nm, 200 nm) | | | | | | | |
| Post-objective illumination power (W/cm$^2$) | 0.65 | 0.0022 – 1.72 **(A-E)** / 0.00045 – 0.65 **(F-H)** | 0.65 | 3.73 | 0.46 | 0.09 | 0.18/0.30 | 0.18 |
| Downsampling along *z*-axis | 2× | | | | | | | |
| Encoding – angular spacing (degree) | 3 | 5 | 7.5 | 3 | 3 | 3 | 3 | 3 |
| Encoding - depth | 7 | 6 | 6 | 7 | 7 | 6 | 7 | 6 |
| MLP – number of layers | 6 | | | | | | | |
| MLP – number of features | 128 | | | | | | | |
| Regularization parameters (TV, RSD) | (1e-9, 5e-4) | | | | | | | |
| Number of iterations (Stage 1 / Stage 2) | 400 / 2000 | 400 / 1000 | 400 / 1000 | 400 / 1000 | 400 / 1000 | 400 / 1000 | 400 / 1000 | 400 / 1000 |
| Duration of Stage 1 (s) | 24.11 | 16.65 | 14.86 | 28.74 | 24.11 | 21.24 | 24.20 | 21.24 |
| Duration of Stage 2 (s) | 193.2 | 80.77 | 76.21 | 152.0 | 96.50 | 90.90 | 96.50 | 90.90 |
| Total duration for reconstruction and estimation (s) | 217.3 | 97.42 | 91.07 | 180.7 | 120.6 | 112.1 | 120.7 | 112.1 |